%% file: main.tex
\documentclass[sigconf, nonacm]{acmart}

\AtBeginDocument{%
  \providecommand\BibTeX{{%
    \normalfont B\kern-0.5em{\scshape i\kern-0.25em b}\kern-0.8em\TeX}}}

\setcopyright{acmcopyright}
\copyrightyear{2023}
\acmYear{2023}
\acmDOI{XXXXXXX.XXXXXXX}

\acmPrice{15.00}
\acmISBN{978-1-4503-XXXX-X/18/06}

\usepackage{pifont}
\usepackage{circledsteps}
\usepackage{listings}
\usepackage{_macros}

\newcommand{\sysname}{Influencer}

\begin{document}

\title[\sysname]{\sysname{}: Empowering Everyday Users in Creating Promotional Posts via AI-infused Exploration and Customization}

\author{Xuye Liu}
\orcid{0000-0001-5876-7229}
\affiliation{
  \institution{University of Waterloo}
  \streetaddress{200 University Ave W}
  \city{Waterloo}
  \state{Ontario}
  \country{Canada}
}
\email{xuye.liu@uwaterloo.ca}

\author{Annie Sun}
\affiliation{
  \institution{University of Waterloo}
  \streetaddress{200 University Ave W}
  \city{Waterloo}
  \state{Ontario}
  \country{Canada}
}
\email{a34sun@uwaterloo.ca}

\author{Pengcheng An}
\affiliation{
  \institution{Southern University of Science and Technology}
  \city{Shenzhen}
  \state{Guangdong}
  \country{Canada}
}
\email{anpc@sustech.edu.cn}

\author{Tengfei Ma}
\affiliation{%
  \institution{Stony Brook University}
  \city{Stony Brook}
  \state{New York}
  \country{United States}}
\email{tengfei.ma@stonybrook.edu}

\author{Jian Zhao}
\orcid{0000-0001-5008-4319}

\affiliation{%
  \institution{University of Waterloo}
  \streetaddress{200 University Ave W}
  \city{Waterloo}
  \state{Ontario}
  \country{Canada}
}
\email{jianzhao@uwaterloo.ca}

\renewcommand{\shortauthors}{Liu, et al.}
\newcommand{\tooltextblock}{\Circled{1}}
\newcommand{\toolimageblock}{\Circled{2}}
\newcommand{\tooluploadimageblock}{\Circled{3}}
\newcommand{\toolpostblock}{\Circled{4}}
\newcommand{\toolimageexplore}{\Circled{5}}
\newcommand{\toolmaskedit}{\Circled{8}}
\newcommand{\tooleditresult}{\Circled{9}}
\newcommand{\toolconcept}{\Circled{6}}
\newcommand{\toolmoreimage}{\Circled{7}}
\newcommand{\toolimgtopost}{\Circled{10}}
\newcommand{\toolregenerate}{\Circled{11}}
\newcommand{\tooltxtoption}{\Circled{12}}
\newcommand{\toolsearchimg}{\Circled{13}}
\newcommand{\toolgeneratecaption}{\Circled{14}}
\newcommand{\tooltxttopost}{\Circled{15}}
\begin{abstract}
Creating promotional posts on social platforms enables everyday users to disseminate their creative outcomes, engage in community exchanges, or generate additional income from micro-businesses. However, creating eye-catching posts combining both original, appealing images and articulate, effective captions can be rather challenging and time-consuming for everyday users who are mostly design novices. 
We propose \sysname{}, an interactive tool to assist novice creators in crafting high-quality promotional post designs, achieving quick design ideation and unencumbered content creation through AI. 
Within \sysname{}, we contribute a multi-dimensional recommendation framework that allows users to intuitively generate new ideas through example-based image and caption recommendation. Further, \sysname{} implements a holistic promotional post design system that supports context-aware image and caption exploration considering brand messages and user-specified design constraints, flexible fusion of various images and captions, and a mind-map-like layout for thinking tracking and post-recording. 
We evaluated \sysname{} with 12 design enthusiasts through an in-lab user study by comparing it to a baseline combining Google Search + Figma. 
Quantitative and qualitative results demonstrate that \sysname{} is effective in assisting design novices to generate ideas as well as creative and diverse promotional posts with user-friendly interaction. 
\end{abstract}



\begin{teaserfigure}
  \includegraphics[width=\linewidth]{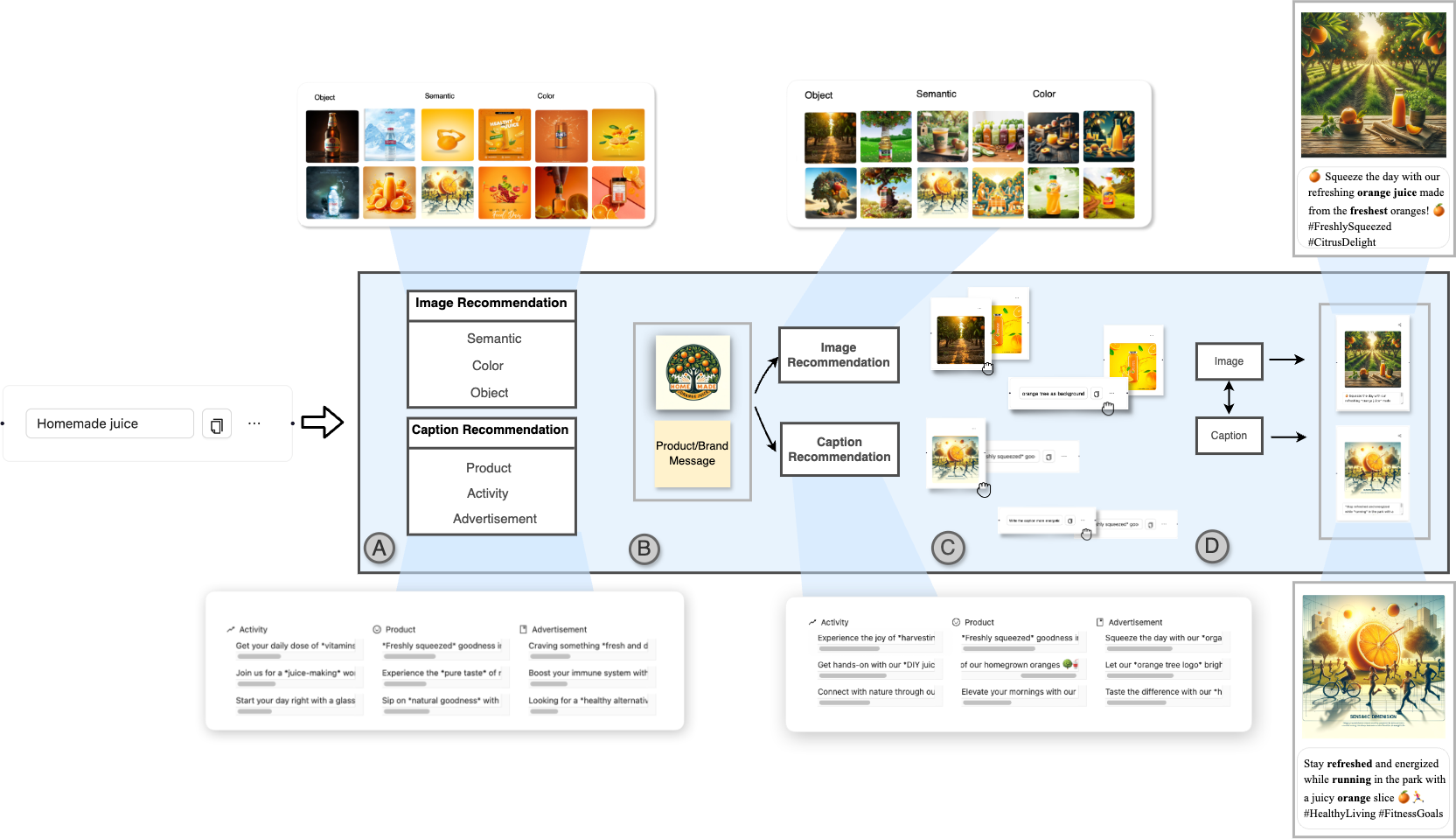}
  \vspace{-6mm}
  \caption{A design novice uses \sysname{} to ideate and make promotional posts to promote her homemade juice. (A) The user starts by input the topic in \sysname{} via a text block and explores the related images and captions in various directions. (B) Context-aware exploration is supported which updates the image and caption recommendation based on a brand/product image or message. (C) Various materials (i.e., image and text) can be flexibly fused to make a new image or caption. (D) \sysname{} allows the user to not only easily create harmonious promotional posts but also quickly obtain multiple post alternatives. }
  \Description{The figure is a pipeline illustrating the promotional post design process in Influencer, which is fully described in the text. }
  \label{fig:teaser}
\end{teaserfigure}



\maketitle

\input{0_Introduction}
\input{1_RelatedWork}
\input{2_SystemDesign}
\input{3_UsageScenario}

\input{4_SystemPipeline}

\input{5_UserEvaluation}
\input{6_Discussion}

\input{8_Conclusion}


\bibliographystyle{ACM-Reference-Format}
\bibliography{main}


\end{document}

%% file: 0_Introduction.tex
\section{Introduction}

Creating promotional posts on social platforms enables everyday users to share their creativity and engage in micro-entrepreneurship, an emerging economic component representing flexible, small-scale business ventures that offer substantial growth potential~\cite{richards2006sustainable}. 
By creating promotional posts on social media, ``micro-entrepreneurs'' could showcase their products or services (such as selling second-hand items, homemade cakes~\cite{metcalf1973starting}), engage with target customers, and build their online presence~\cite{adekunle2020nexus, menon2019grow}.
This helps individuals, such as low-income groups or college students, to generate additional income for 
improving their life qualities, thus contributing to the resilience of the society
~\cite{lessa2023micro}.

Promotional post needs to be crafted with an eye-catching design and packed with informative content to boost brand visibility, engage customers effectively, and ultimately pave the way for higher conversions and sales~\cite{hummell2006synectics}. 
Hence, a successful promotional post requires high-level designer skills. Namely, designers need to iteratively create engaging visuals and effective messages, and thoughtfully blend the two into a harmonious whole.


Therefore, for design novices and everyday users, creating promotional posts is challenging in several aspects. 
First, in the ideation stage, since users often draw inspirations from existing images~\cite{gonccalves2014inspires}, it is challenging for them to create novel alternatives from familiar visual narrations~\cite{truong2006storyboarding}.
Previous research indicates that exploring examples helps users better understand their design direction and facilitates their ideation step from previous creations than starting from scratch~\cite{eckert2000sources, bonnardel2005towards, siangliulue2015toward}. 
Thus, many users rely on existing image search engines (\eg, Google Image Search, Pinterest) to collect examples during the ideation stage; however, this process can be very time-consuming~\cite {siangliulue2015providing}. 
Researchers have developed tools (\eg, MetaMap~\cite{kang2021metamap}, PopBlend~\cite{wang2023popblends}) to support example exploration in the design ideation process, but the search results (design examples) cannot be flexibly edited or modified. 
However, users still need to rely on additional retouching tools (\eg, Adobe Photoshop\footnote{\url{https://www.adobe.com/ca/products/photoshop.html}}, Figma\footnote{\url {https://www.figma.com/}}) to create and refine their designs.

Moreover, 
the coordination between images and captions is essential for creating coherent and persuasive promotional posts that effectively convey the intended brand message and campaign information. 
However, matching the image to the message, or vice versa, can be a challenge in practice, due to the particular themes of the designed post. Without readily usable image resources, users often need to manually create an image to satisfy design constraints and align them with the theme~\cite{horphet2020analysis}. 
Similarly, users need to iterate the caption appropriately to match the image for accurate conveyance of the intended message. 
These design hurdles escalate the threshold of promotional post design for non-professionals. 
While there exists some image editing tools (\eg, Adobe Photoshop) and image captioning tools (\eg, jina\footnote{\url{https://jina.ai/}}), they are either targeted at professional users or lack the flexibility and integrativeness for a streamlined unified workflow: e.g., images and captions need to be created separately and combined manually at a later stage. Furthermore, such a complex, tedious and time-consuming process makes it difficult for novice users to create a sufficient number of design alternatives to compare and evolve their post, which is essential for having a good design outcome.


To address the challenges, we propose \sysname{} (\autoref{fig:teaser}), an AI-empowered interactive tool for users who do not have professional design skills but have needs to create promotional posts in their daily lives (e.g., individual creators, college students, small business operators, product managers, freelancers). 
Drawing design goals from a formative study with five professionals, \sysname{} has been designed to streamline the promotional post design process, featuring four design modules: ideation, context-aware exploration, iterative customization, and delivery of multiple alternatives. 
Specifically, it provides: 1) an example-based multi-dimensional recommendation framework to facilitate the exploration of related seed captions and images to inspire design ideas, 2) a context-aware exploration module that enables users to add complex design constraints including the requirement of matching color schema of brand image or product message, 3) flexible fusion for various design materials via LLMs and Generative Image AI models, and 4) a mind-map layout to organize ideas, track thought process, and present multiple design alternatives. 
To evaluate \sysname{}, we conducted a controlled experiment with 12 design novices comparing our system with a baseline resembling the current workflow (Google + Figma). The results indicate that \sysname{}'s features were appreciated and the system in effective in helping users generate more creative and higher-quality promotional posts.  
In summary, we make the following contributions:

\begin{itemize}
  \item A comprehensive AI-infused pipeline for effective ideation and generation of promotional posts, which integrates multi-dimensional recommendation, context-aware exploration, and conceptual fusion mechanisms, for both images and captions. 
  \item A novel interactive canvas-based tool enabled by the pipeline, \sysname{}, which facilitates users with designing promotional posts and organizing them in a mind-map-like layout to track their thought processes. 
  \item A comparative evaluation of \sysname{} with design novices and gained insight into whether and how the tool is effective in supporting promotional post design.
\end{itemize}

%% file: 1_RelatedWork.tex
\section{Related Work}

\subsection{Example-based Design and Ideation}
Past research indicates that exploring a wide range of design examples enables designers to gain the potential approaches for implementing their ideas~\cite{eckert2000sources}. An effective way to generate novel ideas is to utilize example-based exploration rather than starting from scratch~\cite{siangliulue2015toward}. 
By getting timely inspiration from these examples, designers can generate a lot more ideas than only relying on brainstorming. When individuals heavily rely on existing knowledge and examples before generating new ideas~\cite{jansson1991design}, they may encounter design fixation. 
In addition, simply providing irrelevant inspirations can not effectively address this issue and can even influence design efficiency~\cite{chan2017semantically}.
Hence, it is crucial to maintain a balance between interrelatedness and diversity in example-based recommendations to foster designers' creativity~\cite{perttula2007idea}. 
Psychologists have investigated the relationship between human creativity and association~\cite{gough1976studying, merten1999creativity}. They find that generating creative ideas demands a strong capability for finding an association, rather than only imitating past work~\cite{brown2008guiding}. 
Based on previous examples, experienced designers are more adept at using analogical reasoning to link different concepts than novices~\cite{bonnardel2005towards}.
Thus, novices need more robust support to innovate and combine concepts, which is also essential for promotional post design. 

To facilitate this process, researchers have published free association datasets to simplify the ideation process among the general public~\cite{de2019small}. The small world project\footnote{\url{https://smallworldofwords.org/en/project/home}} is currently the largest dataset for word associations in English with more than 12,000 cue words, supporting users' conceptual level ideation. It demonstrated the importance of facilitating the ideation process at the concept level in promotional post design~\cite{bonnardel2005towards, jansson1991design}. 
Tools have also been proposed to leverage the example-based approach to facilitate ideation in design activities such as generating compound icons~\cite{zhao2020iconate} and visual metaphors~\cite{kang2021metamap}.
Our work is grounded by the findings in the above literature and proposes a novel multi-dimensional ideation method by analyzing the characteristics of images and captions.

\subsection{Image and Caption Recommendations}

Promotional post designs serve as powerful means for marketing and editorial graphics, enabling designers to capture audiences and transmit messages in an engaging and symbolic manner~\cite{khan2021influencer}.
Such designs require the seamless integration of a promotional image and its caption~\cite{tiggemann2022digital}.
Jeong's pioneering work~\cite{jeong2008visual} mentions that posts with visual metaphors are more persuasive compared to posts with literal (non-metaphorical) images.
Therefore, we decide to use visual metaphors as our main image resource in our promotional post design system. We here introduce prior work on recommendations regarding images and captions.

To ease the promotional post design process, various tools have been developed to assist designers in crafting high-quality images.
It is generally required to identify the analogy between two images to have a better chance of finding related images~\cite{kang2021metamap}.
The most common correlation between images is semantic relevance, but an analogy may also involve other topics and categories besides semantic relevance.
Apart from semantic analogy, both color~\cite{kim2017thoughts} and object~\cite{peterson2019aspects} play important roles in establishing metaphorical relationships. Lucero \etal~\cite{lucero2009co} found color distribution creates a specific atmosphere for the audience. 
Prior research has explored the semantic meanings of colors~\cite{jahanian2017colors, kim2017thoughts, wu2018mediating}, aiming to establish links between colors and semantics. Researchers also highlighted the usefulness of color features in representing objects and conveying underlying semantic messages~\cite{kim2017thoughts}. Besides, objects are also important when recommending related images, and designers tend to seek similar elements with similar objects for potential image exploration~\cite{gkiouzepas2011articulating}. 
Therefore, in our work, we seek to use a three-dimensional recommendation framework (semantic, color, and object) for image exploration in the ideation step.

Moreover, prior research has shown that creating interesting and related captions can improve interaction and the sharing of content on promotional post design~\cite{jaakonmaki2017impact, yu2024art}. Traditional caption generation methods~\cite{ren2024product, srivatsan2023text} for posts mainly use natural language techniques to create captions just from the picture, not considering the post's actual use and focusing only on one aspect. 
However, these methods do not fully meet the varied requirements of different types of posts. 
We propose a novel framework that employs a multidimensional strategy for caption recommendation in post design. 
We separate captions in promotional posts into three main topics: activity, product, and advertisement. 
Each of these themes serves a distinct purpose in engaging the audience and driving desired actions. 
Activity-focused promotional posts entail content that centers on engaging actions or events. 
They aim to grab the audience's attention by showcasing active happenings, events, or efforts linked to a brand or group~\cite{schultz2017proposing}. 
Product-related promotional posts seek to convey the attributes, advantages, and value of particular products or services. This marketing strategy aims to attract potential consumers and encourage purchase decisions~\cite{banerjee2009effect}.  Advertisement-focused promotional posts encompass broader marketing and branding initiatives~\cite{kim2021consumer}. They aim to promote the brand as a whole, establish its identity, and convey its unique selling propositions. 
In our work, we incorporate caption recommendation in three dimensions (activity, product, advertisement) in \sysname{} to facilitate the caption exploration in promotional post design.

\subsection{Promotional Post Design Systems}

There are a variety of off-the-shelf tools that support post design with different focuses, such as professional image editing (e.g., Adobe Photoshop), collaborative UI design (e.g., Figma), and user-friendly graphic design (e.g., Canva). While all these tools provide powerful post image editing features from template-level to object-level, they have a steep learning curve for novice designers and require much manual user effort. Thus, the research community has been exploring AI-driven technologies for design assistance. 


In particular to the design of promotional posts, customization techniques are widely explored to facilitate and accelerate promotional post design. 
Most of the existing research has primarily focused on image customization in the post design scenario. 
Common image customization techniques include masking and editing specific areas, generating new images based on a similar prompt context, and regenerating parts of images for variations~\cite{ramesh2021zeroshot}. 
More recent works include AnyDoor~\cite{chen2023anydoor} which used a diffusion-based deep learning method to support object-level image customization. Some works also try to incorporate text or image concepts to customize target images. 
For example, Chen \etal~\cite{chen2023photoverse} enabled various prompt inputs by incorporating a dual-branch conditioning mechanism to customize images in different concepts. 
Further, Kumari \etal~\cite{Kumari_2023_CVPR} supported composing multiple concepts to generate images based on the given texts.
On the other hand, beyond image customization, the generation and customization of captions for promotional posts also facilitate the promotional post design process. 
Large Language Models (LLM) including ChatGPT~\cite{openai_gpt-4_2023}, Llama~\cite{touvron2023llama}, and Claude~\cite{anthropic_model_card_2023} have been proposed to customize caption by rewriting the prompt based on user's need. 
More recent studies have used Context Sequence Memory Network to customize descriptive captions and predict hashtags based on the query images~\cite{Park_2017_CVPR, park2018towards}. 
In this work, we integrate multiple customization techniques such as regenerating images with similar prompt context and mask editing images into our tool to improve the promotional image design process. 
We also consider a simplified and practical interaction technique for promotional post customization, enabling users to conceptually customize captions and images through drag-and-drop selected images or texts as context prompts on a mind map.



Additionally, researchers have proposed various intelligent tools to help novices automate parts of the design workflow. 
For example, there has been a variety of work in domain-specific applications with compound design tasks. Yin \etal~\cite{10.1145/2502081.2502116} developed a system for automatically generating magazine-like visual summaries from traditional social media posts for efficient mobile browsing. 
Qiang \etal~\cite{qiang2017learning} introduced a graphical model that learns to generate scientific posters from research papers. 
Tyagi \etal~\cite{Tyagi_2022} proposed a tool to generate infographics from a user sketch. 
Shi \etal~\cite{10.1145/3544548.3581070} developed a holistic color authoring system that supports 2D palette extraction, theme-aware, and spatial-sensitive color. 
Zhao \etal~\cite{zhao2020iconate} created an automated system that generates icon suggestions by considering semantics, style, and space to support compound icon design.
Recent work in aiding creative design~\cite{bae2020spinneret} also adopts mind-maps to mimic the process of making associations among design materials. 
However, all of the above tools mainly focus on the ideation step of design, requiring designers to switch to other tools for the actual design work. Also, none of the above tools accommodates user-specified prompt or context images (e.g. brand image) based on design constraints, which are essential to maintain consistency and effectively communicate with the target audience.

%% file: 2_SystemDesign.tex
\section{Formative Study}

Our primary target audience is design novices such as product managers, marketers, small business operators, freelance creators, or anyone who needs to quickly make attractive promotional posts without professional training. 
To better understand their needs, we conducted a formative study, and here we describe its setup and yielded design goals that have guided the development of \sysname{}.


\subsection{Setup}

We recruited five professionals (two women and three men, aged 20-30), including two design researchers in the IT company, a marketer experienced in designing promotional posts, and two software engineers who collaborate with product managers, marketers, and designers on product promotion. 
The study included a 30-minute semi-structured interview with each participant. 
The interview questions covered how they would normally design promotional posts, including how they got their inspiration, and how they would craft and iterate the posts. They were also asked to identify difficulties they encountered throughout the design process and raise their needs for support. 
The interviews were audio recorded and then transcribed for further analysis. All the authors used the open-coding method to independently conduct a thematic analysis of the interview transcripts and notes. Then, all the authors reviewed and synthesized the results to ensure a comprehensive understanding of the findings.

\subsection{Formulating Design Goals}

Based on our formative study, we distill the following design goals to inform our development of \sysname. We refer to the participants in our interviews as E\#.

\subsubsection{\textbf{R1: Help explore images and captions in diverse dimensions}}
Designers create new ideas through association and recombination from previous examples~\cite{bonnardel2005towards}. Traditional methods to collect inspiring examples of images and captions for the ideation stage include searching online resources from search engines (e.g., Google Search), which is usually cumbersome and time-consuming. The participants confirmed that they all seek inspiration from others' works, including content, color schema, style, and composition. Exposure to rich information often sparks new ideas~\cite{leonard1997spark}. However, when talking about searching exemplar images and captions for inspiration, two design novices (E1, E3) reported that they did not have effective techniques for using search engines for this purpose. Four participants (E1-2, E4-5) suggested a diverse example-based image recommendation would be desirable in this task. 

\textbf{R1.1: Image recommendation.} 
Image recommendation plays a pivotal role in inspiring visual design for promotional posts by offering a wide array of imagery to draw inspiration. However, traditional methods lack multi-dimensional exploration of the image beyond basic keywords or tags, overlooking crucial aspects such as object, content, and semantic meaning. E1 mentioned: \qt{Searching images on the search engine results in many irrelevant or useless images. I often spend a lot of time on image selection.} 
Moreover, creativity researchers~\cite{kang2021metamap} demonstrated that multi-dimensional image recommendation can help explore visual metaphors including semantics, color, and shape, which was confirmed by E2. 
In addition to images with similar semantics and color, designers desire images that share similar objects. However, they are not particularly inclined towards images with similar shapes. 
E3 commented: \qt{I find the images with similar objects more useful than just similar shapes. It's about making that connection and conveying the right message in our promotional stuff. Shapes alone don't tell the story.} The participants' feedback suggests that an example-based multi-dimensional recommendations (\ie, semantic, color, and object) should be incorporated into the image recommendation module. 

\textbf{R1.2: Caption recommendation.}
The caption in a promotional post is greatly determined by its design intention and usage scenarios~\cite{jaakonmaki2017impact}. For instance, E4 said, \qt{In a promotional post, captions should set the scene, provide context, and align with the campaign's message. It's about understanding that a caption serves multiple roles.} 
E5 expressed frustrations towards the traditional approach to caption recommendation and creation for promotional posts: \qt{Simple captions aren't enough to make our products stand out, engage our audience, and tell a great ad story; we need captions that truly speak to every aspect and make our content stand out.} Considering the influence of context and its role in shaping the captions in the post, a system should establish a multi-dimensional understanding of the design intention (such as the product, activity, and advertisement of the promotion). Thus, multi-dimensional caption recommendations could ease the users' creative process.

\subsubsection{\textbf{R2: Support context-aware exploration considering brand messages and user-specified design constraints}}\label{sec:R2}

Complex post design requirements can be overwhelming for design novices~\cite{evans2004domain}. Existing off-the-shelf images and caption recommendation tools can offer some inspiration to users~\cite{feng2012automatic, choi2023creativeconnect, gal2022image, handayani2015instagram}. However, there are two unaddressed opportunities: 1) understanding the semantics of the image, and 2) accommodating user-specified design constraints, such as color schemes and related product or branding requirements. Although it was shown in R1 that using a multi-dimensional recommendation method can effectively inspire design novices with new ideas, such a method falls short when designers have specific design constraints such as requiring the right color schema to fit with the brand or product image or making sure captions fit the brand's values. Thus, a more advanced image and caption recommendation/modification mechanism is needed for iterative ideation. 


\textbf{R2.1: Image-aware.} 
When given specific images for material exploration in promotional post design, it is essential to engage in iterative ideation to ensure the exploration aligns with these images. 
E2 confirmed that \qt{It's crucial to update material exploration from multiple dimensions like semantics, color schemes, objects, and usage scenarios based on the given images. This can save time in searching for materials that are compatible with the provided images.} Thus, if given related images, context-aware exploration should be conducted, ensuring that the exploration is updated to align with these images.


\textbf{R2.2: Prompt-aware.} 
A promotional post design often has different demands based on different goals, such as advertising, entertaining, and informing~\cite{batra2016integrating}. Designing different prompts based on varying needs to update material exploration may help designers engage in more directed design efforts. E4 emphasized this requirement in our formative study: \qt{Using prompts efficiently helps narrow down the search to exactly what you need, so you don't have to start your search over for different requirements.} 
Therefore, material exploration for promotional post design needs prompts to support iterative ideation by conveying user-specific demands. 

\subsubsection{\textbf{R3: Flexible fusion of various images and captions to customize designs}}

Four participants suggested that it would be efficient to conceptually fuse different elements via dragging and dropping. \qt{It's like using one element as a prompt background to customize another element on the canvas, making it easy for designers to create prompts and upload images to customize the target image or caption through conceptual fusion.} 
Users have various preferences and expectations in different scenarios of the promotional post design processes. Hence, it is necessary to provide users with the right amount of customization in the mind-map interface. The tool should make customization easier by automating tedious steps and offering creative fusion options so that they can rapidly enable design iterations and deliver design alternatives. According to our formative study, we identified the following key aspects related to material fusion.

\textbf{R3.1: Image-caption fusion.} 
In promotional post design, an artful fusion of images with precisely crafted text messages is essential. This dynamic combination allows for the creation of a compelling narrative, enhancing audience engagement and guiding their perception of the product or message being promoted. As E5 mentioned: \qt{Pairing the right caption or prompt with a thoughtfully chosen image can create marketing magic.}

\textbf{R3.2: Image-image fusion.} 
Strategically integrating a variety of images is a powerful technique for crafting visually stunning promotional posts. By combining different images, a story can be woven, emotions can be evoked, or a product's unique features can be effectively highlighted. E2 emphasized: \qt{Images engaging in a visual dialogue on the canvas is key. It's about selecting images that complement each other, conveying a unified message to the audience.}

\textbf{R3.3: Prompt-caption fusion.} 
The collaboration of a well-crafted prompt with a compelling caption yields an impactful promotional post. The prompt acts as a guide, steering the direction of the caption and ensuring it aligns with the intended message or brand voice. E1 echoed, \qt{The prompt is like a catalyst for crafting the perfect caption. It triggers creativity and ensures the caption hits the mark, captivating attention and driving the desired response from the audience.} 

\textbf{R3.4: Prompt-image fusion.} 
Many design novices struggle with how to edit images to achieve their desired outcomes. They often lack design expertise and are not proficient with effective image customization with tools. As E2 mentioned: \qt{Design novices like us don't really know how to proceed with the editing to get the desired effect. If I want to change the background or the main color of one image, I don't know how to use related design tools to implement it.} Hence, the tool should allow users to employ the prompt to customize the image by easily ataching the text prompt to the target image.

\subsubsection{\textbf{R4: Design a mind-map layout to organize ideas and track the thought process.}}

According to our interview, designers can easily get lost in iterative exploration of images and captions during ideation without tracking their exploration path. 
Canvas lets designers freely create by allowing them to make, move, track, and link elements anywhere~\cite{staiano2022designing}.
Previous work also indicates that tracking the thinking path with mind-maps during exploration is important for ideation~\cite{dong2021promoting}. E3 suggested that \qt{mind-map is one way to connect and recombine materials to generate new ideas.}
Maintaining a thorough history of their creative process is a crucial design principle that aids in fostering creativity. 
E1, E2, and E4 suggested that \qt{When updating material exploration by uploading images or creating prompt texts, using links to show that an update has been made. It's like marking our trail, so we know where we've been. 
The same goes for the flexible blending feature; a link means we modified the target material using flexible blending.
} 
Thus, the tool should support thought process tracking according to users' design actions.

%% file: 3_UsageScenario.tex
\section{\sysname{} Design Overview}

\begin{figure}[tb]
  \centering
  \includegraphics[width=\linewidth]{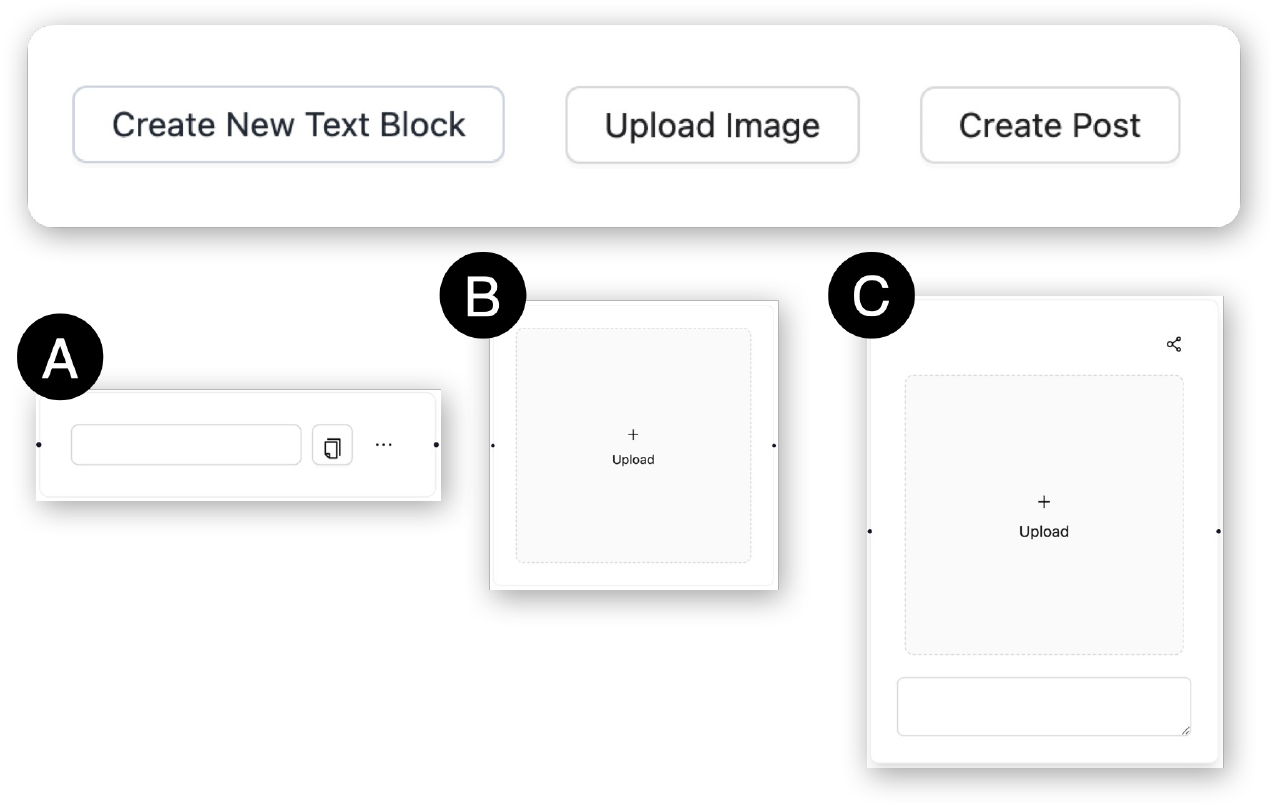} 
  \vspace{-9mm}
  \caption{\sysname{} provides three basic types of interactive blocks on its mind-map canvas. Clicking \textit{Create New Text Block} adds (A) an empty text block. Clicking \textit{Upload Image} generates (B) an empty image block. Clicking \textit{Create Post} produces (C) a blank post blockthat can fuse selected images and captions.} 
\label{fig:block}
\Description{} 
\end{figure}

\begin{figure*}[tb]
  \centering
  \includegraphics[width=\linewidth]{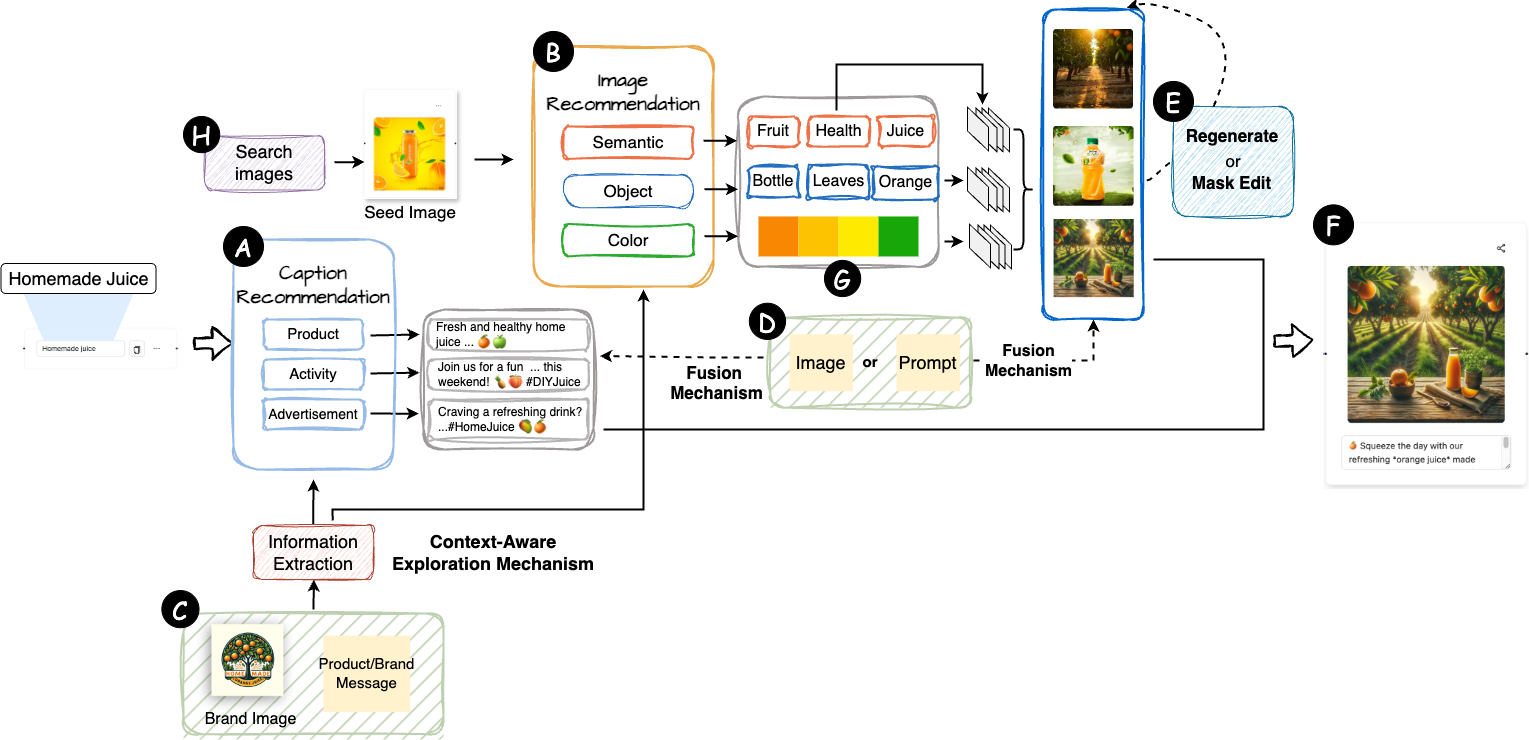} 
  \vspace{-7mm}
  \caption{System architecture and workflow of \sysname{}. From an input topic, the system first analyzes the input to return searched images (H) and the user can select a seed image from the search results. At the same time, the Caption Recommendation (A) digests the input topic and recommends captions from three dimensions (i.e., product, activity, advertisement). Next, \sysname{} can conduct image recommendation (B) based on the seed image in three dimensions (i.e., semantic, object, color) and suggest images for each dimension (G). It also supports context-aware exploration (C) based on a brand image or product/brand message to find materials aligning with the brand. Besides, \sysname{} allows users to customize the images via Regenerate or Mast Edit (E) and conceptually fuse design materials (i.e., images and captions) with an image or prompt reference (D). Together, it generates an aesthetically pleasing promotional post design with harmonious content. A user can further iteratively explore more design recommendations (A, B, G) and generate multiple design alternatives to share on social media (F).} 
\label{fig:pipeline}
\Description{} 
\end{figure*}

\begin{figure*}[tb]
  \centering
  \includegraphics[width=\linewidth]{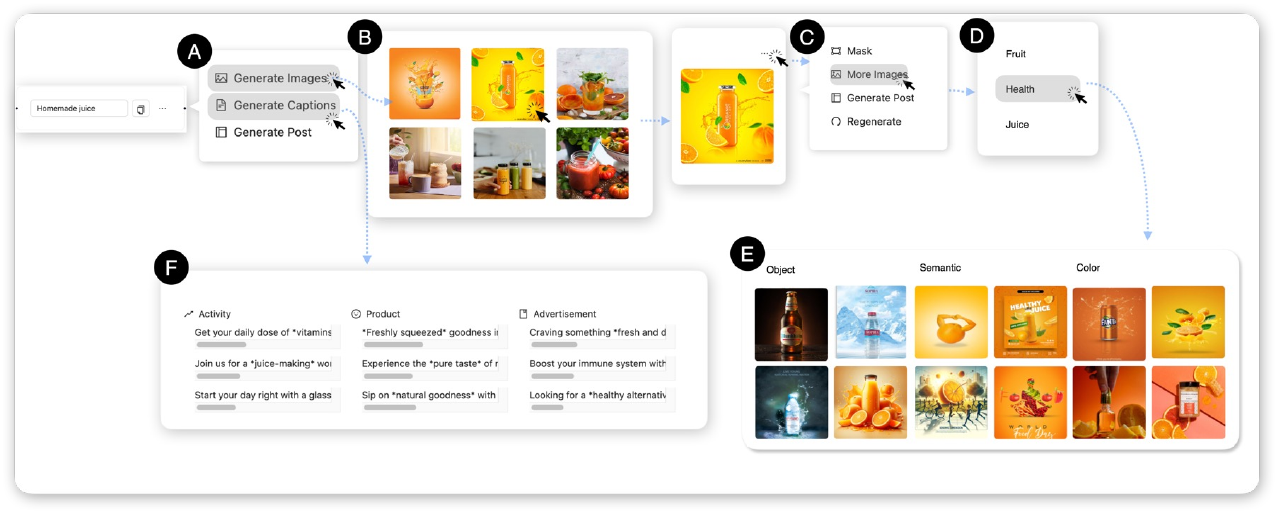} 
  \vspace{-9mm}
  \caption{Ideation with \sysname{}: (A) From a text block, a panel providing three options (\textit{Generate Images}, \textit{Generate Captions}, \textit{Generate Post}) for users to ideate captions or images and generate posts; (B) An assembly of image search results based on users' input topic; (C) From an image block, a panel providing four options for users to customize (\textit{Regenerate}, \textit{Mask}), conduct image recommendation (\textit{More Images}), or produce post directly (\textit{Generate Post}); (D) A panel providing related semantic keywords to help users explore diverse materials; (E) An image recommendation block displaying results in three dimensions (i.e., semantic, object, color); (F) A caption recommendation block displaying results in three dimensions (i.e., product, activity, and advertisement).}
\label{fig:ideation}
\Description{} 
\end{figure*}

\begin{figure*}[tb]
  \centering
  \includegraphics[width=\linewidth]{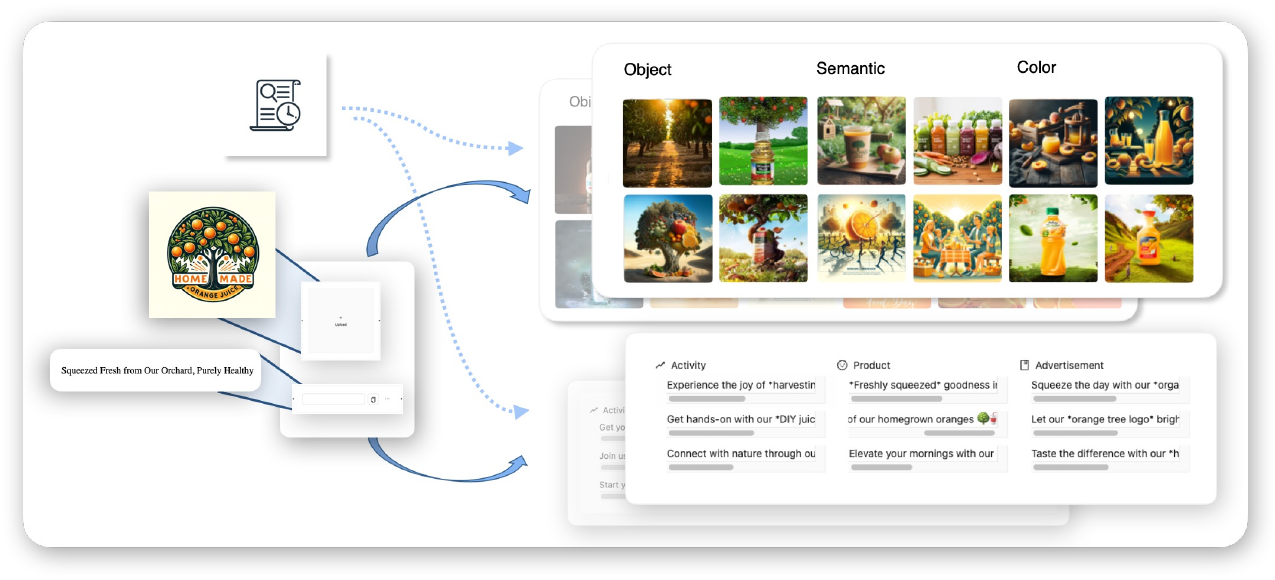} 
  \vspace{-9mm}
  \caption{Context-Aware exploration with \sysname{}: (A) Uploading a brand image and dragging to update the target image recommendation block and caption recommendation block; (B) Creating a text block with product message and dragging to update the target image recommendation block and caption recommendation block; (C) Updated image recommendation block; (D) Updated caption recommendation block.} 
\label{fig:context}
\Description{} 
\end{figure*}

\begin{figure*}[tb]
  \centering
  \includegraphics[width=\linewidth]{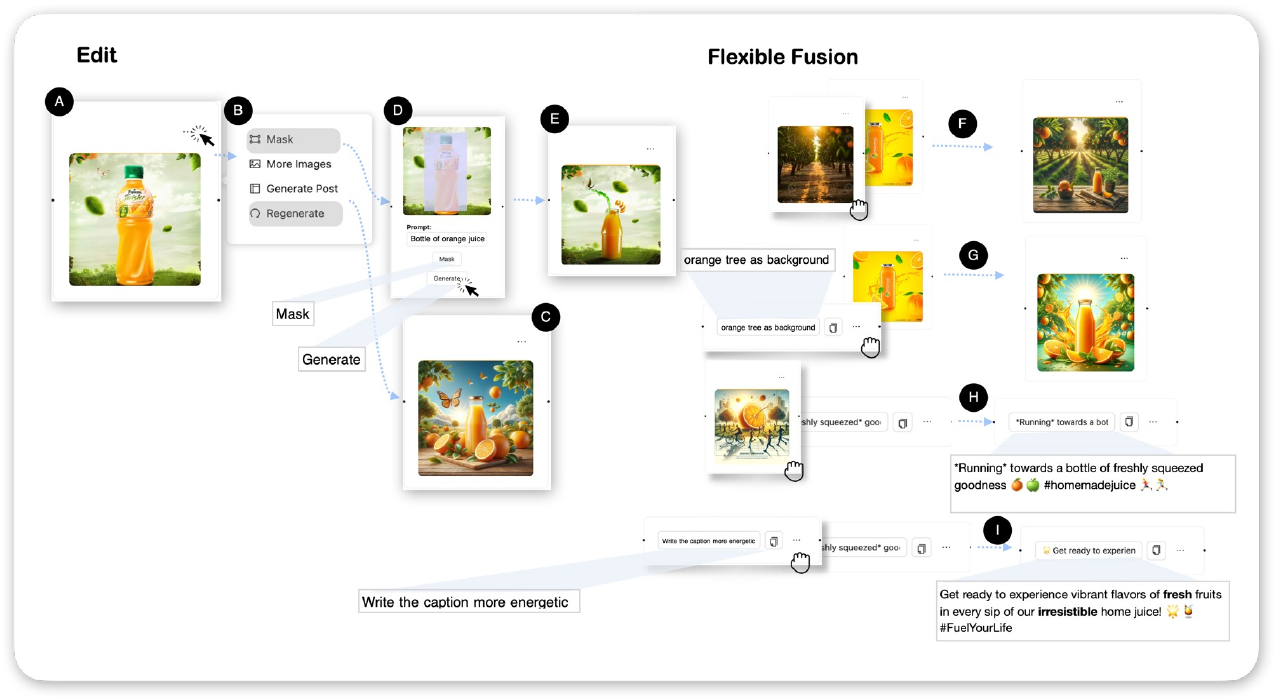} 
  \vspace{-9mm}
  \caption{Customization with \sysname{}: In the \textit{Edit} mode, an image (A) can be modified with ``Mask'' and ``Regenerate'' for customization (B). ``Regenerate'' produces a semantically similar new image (C). ``Mask'' allows for framing the part to customize with a prompt (D) and thus produces a new image (E).  
  In the \textit{Fusion} mode, images and/or text can be freely combined by dragging an image or text block with a prompt to the target image or caption (F, G, H, I).
  } 
\label{fig:customize}
\Description{} 
\end{figure*}

\begin{figure}[tb]
  \centering
  \includegraphics[width=\linewidth]{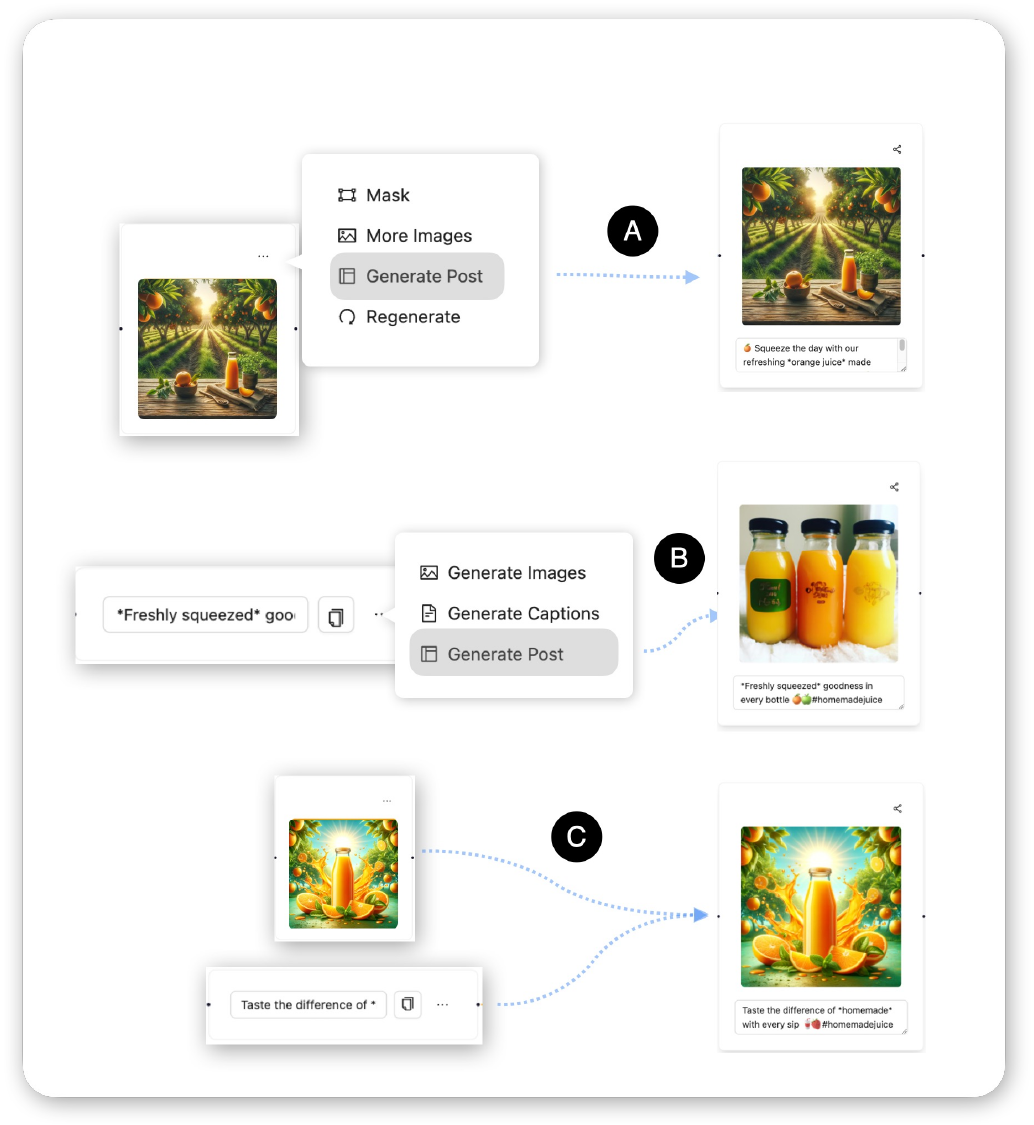} 
  \vspace{-9mm}
  \caption{Post generation: A post can be generated based on a selected image (A) or caption (B) by clicking \textit{Generate Post}. Alternatively, a blank post block can be created to combine the selected image and caption (C).} 
\label{fig:post}
\Description{} 
\end{figure}

Based on the aforementioned design goals, we developed \sysname{}, an interactive, AI-empowered, and canvas-based tool that helps design novices in four main steps of promotional post design (\ie, ideation, context-aware exploration, customization, and post generation), all through a set of simple interactive blocks (\autoref{fig:block}).
\autoref{fig:pipeline} shows an overview of the system backend and workflow, which consists of five main modules: (A) \textit{Caption Recommendation}, (B) \textit{Image Recommendation}, (C) \textit{Context-Aware Exploration Mechanism}, (D) \textit{Flexible Fusion Mechanism}, and (E) \textit{Image Regenerate and Mast Edit Module}. 

In particular, as shown in \autoref{fig:ideation}, users' material (image and caption) searching process is supported through related concepts based on their search input. After identifying an image or caption of interest, users can further explore the recommendation images based on three features: semantics, color, and object, and explore the recommendation captions in three contexts: product, activity, and advertisement (\textbf{R1}). As shown in \autoref{fig:context}, users can conduct context-aware exploration for captions and images by dragging brand messages (\ie, brand image, product message, and brand value) to the caption and image recommendation block (\textbf{R2}). Users can further contextually customize design materials by fusing any images and captions during the promotional post design, as shown in \autoref{fig:customize} (\textbf{R3}). All the activities are conducted on a mind-map which records the thinking path generated by users for a quick recollection (\textbf{R4}).

\subsection{User Interface and Blocks}
The user interface of \sysname{} is based on an interactive canvas allowing users to create three types of blocks: text block (\autoref{fig:block}.A), image block (\autoref{fig:block}.B), and post block (\autoref{fig:block}.C), as well as offering two types of recommendation blocks: image and caption. 

The image block provides four options for users to further explore, customize, or utilize images in their promotional post design: 1) ``Regenerate Image'' (\autoref{fig:customize}.B) to produce a similar image with the same context, 
2) ``More Images''(\autoref{fig:customize}.B) to browse recommended images based on the selected image in three dimensions (\ie, semantics, color, object) in a image recommendation block (\autoref{fig:ideation}.E), which can further obtain images along the branches (\autoref{fig:ideation}.C), 
3) ``Mask Edit'' (\autoref{fig:customize}.B) to edit an image by drawing a mask and providing a prompt for generating a new image, 
and 4) ``Generate Post''(\autoref{fig:post}.A) to craft a post in a post block directly based on the current image.
To minimize user interaction for efficiency, \sysname{} apply every user selected or uploaded images with those four options to streamline any design activities (\ie, ideation, context-aware exploration, customization, and post generation) from the block. 

The text block provides three options (\autoref{fig:ideation}.A) for users to ideate captions or images and generate post during the design process: 
1) ``Generate Images'' to obtain an assembly of image search results based on the given topic, 
2) ``Generate Captions'' to explore recommended captions in three context dimensions (\ie, product, activity, advertisement) in a caption recommendation block (\autoref{fig:ideation}.F), which can keep expanding along any branch, 
and 3) ``Generate Post'' to directly produce the post based on their input text in a post block (\autoref{fig:post}.B). 

In \sysname{}, images or text can be as context to help users do further exploration. Users can create a text block writing brand message or the requirement of the promotional post design and drag it to the image (\autoref{fig:ideation}.E) or caption recommendation block (\autoref{fig:ideation}.F). In this way, users can get new image or caption recommendation aligning with the prompt written in the text block. Also, users can create an image block (\autoref{fig:block}) and upload the context image like brand image and drag it to the image (\autoref{fig:ideation}.E) or caption recommendation block (\autoref{fig:ideation}.F). Then user will get a new image and caption recommendation block aligning with his uploaded image with the similar semantic meaning and color schema. At the same time, users can also drag a text block or an image block and connect it to any target image block, text block for conceptual customization.
Such customization and context-aware exploration are shown in orange arrows to differentiate from the users' initial exploration paths that are shown in blue arrows.
Furthermore, users have the option to not only generate posts based on a single image or caption but also select their preferred image and caption directly on the canvas and fuse them to a post block for production.

\subsection{Usage Scenario}

Before diving into our approach and details, we explain how the design novices use \sysname{} via a simple usage scenario. 
Suppose Crystal is a micro-entrepreneur who owns a small grocery store, and she needs to promote a new product, homemade orange juice. 

Crystal launches \sysname{}. She first creates a Text Block (\autoref{fig:block}.A) on the canvas and wants to explore images related to her promotional topic: ``homemade juice.'' 
She clicks the ``...'' button that shows a list of exploration and post generation options (\autoref{fig:ideation}.A).
From there, she clicks the ``Generate Images'' button (\autoref{fig:ideation}.A) to create a search image block (B), consisting of a set of images related to the given topic. 
After identifying an image of interest, she further explores the recommendation images based on 3-dimensional features (semantics, color, and object) by clicking on the ``More Image'' button (\autoref{fig:ideation}.C). After clicking this button, \sysname{} requests her to select a related semantic keyword from the semantic keyword panel (\autoref{fig:ideation}.D) (\textbf{R1}). Crystal wants the focus of the promotion to be on health. Then she click Health from (\autoref{fig:ideation}.D) and create an image recommendation block (\autoref{fig:ideation}.E) for more images in health semantic meaning.

Crystal looks at the images from the image recommendation block and finds a satisfied image in (\autoref{fig:customize}) but thinks \qt{The image's meaning is closely related to my topic, but I need more alternatives with similar meaning and want to change the type of juice shown.}
Next, she click the ``Regenerate'' (\autoref{fig:customize}.B) to generate an image with the same semantic meaning (\autoref{fig:customize}.C). She also wants to try to click ``Mask Edit'' button (\autoref{fig:customize}.D) to get more alternatives. After clicking ``Mask Edit" button, she clicks ``Mask" button from (\autoref{fig:customize}.D) to frame out the mask she wants to replace and writes the prompt in (\autoref{fig:customize}.D). Then she clicks the ``Generate" button to generate a new image in (\autoref{fig:customize}.E). 

At the same time, Crystal wants to further explore other alternative paths of design, since she does not have sufficient design training and wants to see if there are better options. Then she tries another path to explore useful captions for her promotional post. 
She thus clicks the ``Generate Captions'' button (\autoref{fig:ideation}.A) to obtain a caption recommendation block (\autoref{fig:ideation}.F).
She checks this caption recommendation block and finds one caption suitable for her post. 
She clicks this caption which generates a text block including this caption and edits the text based on her preference (\autoref{fig:post}.C). Then, she directly generates a post by clicking the ``Generate Post'' button (\autoref{fig:post}.C).
Through all the above actions, she creates multiple post alternatives for her design and she is quite satisfied (\autoref{fig:post}). 

However, she thinks the promotional image needs to align with the brand image (\autoref{fig:context}.A) of her grocery and some new descriptions of the promoted product (\autoref{fig:context}.B). 
Using brand images or new descriptions directly to find materials may ignore previous searches and context, potentially leading to recommendations that do not match the original intent and needs. 
Therefore, Crystal implements a context-aware exploration to find more materials for post design based on the previous search. She links the brand images and the text block with new product descriptions to the caption and image recommendation block. 
\sysname{} returns newly recommended images and captions in the image (\autoref{fig:customize}.C) and caption recommendation block (\autoref{fig:context}.D) from multiple dimensions which align with the color scheme, object, and semantic meaning of the brand image and the new product descriptions. 
\qt{Nice! I got new materials for the post design!} Crystal and feels happy about the efficiency. 
Using similar interactions above, she continues to customize and generate some alternative designs (\autoref{fig:post}.ABC) (\textbf{R2}).

Crystal is satisfied with the content of the current image design but feels styles are less appealing when compared with another image despite not liking its content(e.g., the left two images in\autoref{fig:customize}.F). 
She wants to combine these two images; thus she connects this image as a background to that seed image for conceptual fusion (\autoref{fig:customize}.F).  
A new image is then generated with a style similar to the background image and the same semantic meaning as the seed image (\textbf{R3}). 
\qt{Looks better!} she thinks. 
If there is no satisfactory image on the canvas, she can also create a text block with the style description and drag it to the seed image for the above conceptual fusion. For example, she drags a text block with the text of ``orange tree as background'' to the seed image, \sysname{} generates a new image featuring an orange tree background while retaining the orange juice bottle from the seed image (\autoref{fig:customize}.G). 
Besides, \sysname{} supports conceptual fusion on captions by dragging existing images or text blocks with prompts to customize target captions (\autoref{fig:customize}.HI). 
Thus, Crystal drags a text block with a prompt suggesting ``write the caption more energetic'' in the target caption (\autoref{fig:customize}.I). She takes a look at the customized caption: \qt{Awesome! That was quick. This caption is exactly what I needed!} 

During the process, Crystal finds thinking path very useful. When she get lost in the promotional post design process, the tracked thinking path can remind her of the previous exploration and customization (\textbf{R4}). 



%% file: 4_SystemPipeline.tex
\section{\sysname{} System}
In this section, we describe the backend of the \sysname{} system in detail, particularly on the implementation of the multi-dimensional recommendation, context-aware exploration, and multiple fusion mechanisms (\autoref{fig:pipeline}).


\subsection{Dataset Building}

Previous work emphasizes the importance of using associations from previous examples to create new ideas~\cite{wilkenfeld2001similarity, bonnardel2005towards}. Besides, based on previous work and our formative study results, we find that the creation of promotional images involves analogy between all kinds of information~\cite{carroll1994visual}, including semantic~\cite{petridis2019human, phillips2004beyond}, color\cite{kim2017thoughts}, and object~\cite{gkiouzepas2011articulating}. To build the post image dataset with reasonable associations between each example, we performed various preprocessing steps to prepare our dataset. 
Specifically, we use Small World Dataset~\cite{de2019small} to get a keyword association dataset and collect a Post Image dataset following Kang \etal~\cite{kang2021metamap}'s approach with further refinement.


\textbf{Key Association dataset.} 
The Small World Dataset~\cite{de2019small} is the largest English free word association resource, containing over 12,000 cue words. This dataset captures how people remember and recall concepts through word associations~\cite{ma2013evocation}. It includes responses from 100 participants who provided three associated words for each cue, leading to a comprehensive collection of word associations. Then association strengths were calculated based on the frequency of each response relative to the total number of responses~\cite{cattle2017predicting}. We then formed the keyword association dataset as a directed graph, with forward associations (from responses to cues) and association strengths as weights. It includes words within two association distances from the topic words, creating a dataset with 7,407 words.


\textbf{Post Image Collection.} Inspired by \cite{kang2021metamap}, we used Pinterest as our main image source for novice designers since it is the most popular website for them to find image examples. Kang et al.~\cite{kang2021metamap} built a keyword-image dataset utilizing the keyword association dataset collected before with 4,861 descriptive words and 76,686 images where images mainly focus on the advertising creative images. Since advertising creative images are a key source for promotional post design~\cite{vanolo2008image}, we updated the dataset based on feedback from our formative study to include object dimension which is more valued by users. This helps design novices have a clearer idea and examples of how to create promotional posts. To identify related objects within our dataset, we fine-tuned the MMDetection model using SAHI~\cite{laurer2022less} and applied it to each image in the dataset sourced from \cite{kang2021metamap}. The image-object relationship tables were added to the dataset to help designers to explore.

\subsection{Image and Caption Recommendation}

\subsubsection{\textbf{Image Recommendation}} \sysname{} recommends images from three dimensions (i.e., semantics, color, and object) to provide related and diverse examples, as detailed below. 
Therefore, we extract the corresponding features from a seed image and get the top-ranked images. Due to space limitations of the block, we display the top four images in each recommended dimension and allow designers to scroll left and right to view more images.

\textbf{(a) Semantic.} We propose distinct methods for semantic image recommendation based on two types of images: those available in the dataset and those uploaded by users. For images within the dataset, we have already identified highly related concept words through the computation of concreteness and imageability scores (ranging from 0 to 1) sourced from \cite{liu2014automatic}, collected through human annotation and synonym expansion. Our approach prioritizes words associated strongly with perceptible concepts and evoking mental imagery. According to our formative study, designers mentioned the importance of incorporating user-uploaded images to enrich the exploration of design materials in promotional post design. To address this, we leverage the state-of-the-art image captioning model in framework OFA~\cite{wang2022ofa} to extract key information from the uploaded image. Additionally, we fine-tune the cutting-edge text classification models mDeBERTa-v3~\cite{laurer2022less} to identify the highly related concept words corresponding to the uploaded image. After getting the highly related concepts, the top four images are randomly selected from the qualified candidates. By doing so, we increase the diversity of the returned images when the semantic correlation remains the same. 

\textbf{(b) Color.} Our approach involves suggesting images with analogous colors by analyzing the dominant color from the image. ColorThief\footnote{https://github.com/lokesh/color-thief} is adopted to retrieve the main theme color of the image. To ensure a speedy search, we search images from related concepts.

\textbf{(c) Object.} \sysname{} recommends images with similar objects based on object detection. To facilitate the search speed and make sure that the image has some semantic relationship with the original image, we only search images in the related concepts. MMDetection models leveraging SAHI~\cite{Akyon_2022} is used to extract the main object and retrieve the top probable items from the image. The top four images with similar objects are then recommended. The reason of using the SAHI framework is that there are many small but important objects in the promotional image based on our interview.

\subsubsection{\textbf{Caption Recommendation}} 
Addtionally, \sysname{} recommends captions from three dimensions (product, activity, advertisement) to provide related and diverse examples. Specifically, activity-focused posts emphasize engaging actions or events, product-focused posts describe the features and benefits of goods or services, and advertisement-focused posts address broader marketing and branding efforts.
Based on the analysis of promotional posts on Instagram~\cite{chu2017vaping}, there are five themes in total including product, activity, advertisement, text, and others. Based on their statistic of categories and the results of our formative study, we recommend our captions in three contexts (i.e., product, activity, and advertisement). We use GPT-4 to generate caption recommendations based on the user input prompt $T_p$ with the following prompt:
\lstset{
      basicstyle=\ttfamily\footnotesize,
      columns=fullflexible,
      frame=single,
      breaklines=true,
      breakindent=0pt
    }
    \begin{lstlisting}
    Prompt: Generate three promotional captions for each dimension (product, activity, advertisement) based on the given text: <T_p>. please also highlight the keywords with asterisks and keep rendering icons in each caption. 
    \end{lstlisting}

\subsection{Context-Aware Exploration}

In addition to image/caption recommendation from scratch, \sysname{} further supports context-aware exploration which is built upon the image or text block dragged by the front-end. Since this task is to help designers update their exploration results for images or captions based on the provided materials (\ie, brand image, product message, etc.), where the goal is to learn the mapping from provided materials to target exploration block. As we stated in Sec. \ref{sec:R2}, existing off-the-shell recommendation tools cannot fulfill our customization requirements. Our goal is to recommend new images or captions reflecting the user-specified theme and design constraints extracted from provided materials, while most existing models are trained only to recognize general features without distinguishing specific dimensions for each image or text, nor considering the context of previous searches and selected seed images or captions. Based on the findings of our formative study, users find it useful for recording their explored history. The purpose of this feature is to update the image or caption recommendation based on the given context, created from the seed images $I_s$ and seed text $T_s$. Also, we refer the image context as $I_c$ and the text context as $T_c$. Four main tasks should be considered to implement:

\textbf{(a) Text Context-Aware Exploration of Images.} To achieve text context-aware exploration of images, we utilize text prompts $T_c$ to update the image recommendations block. For instance, users can modify the recommendation context, such as searching for similar images with different colors or alternate objects, ensuring a diverse and visually appealing array of options. To ensure we can update the image recommendation reasonably, we first summarize the keyword list into those three dimensions (semantic, color, object) in the collected dataset. Then, we use an advanced image captioning model ofa\_image\-caption\_coco\_large\_en~\cite{wang2022ofa} to extract the image description $D_i$ from seed image $I_s$ and combine it with the text context $T_c$ to form a new contextual prompt $P_t^I$ = $D_i$ + under the context of +  $T_c$. 
Next, we run the text classification models mDeBERTa-v3~\cite{laurer2022less} to categorize the contextual prompt $P_t^I$ in three dimensions (\textit{e.g.,} semantic, object, color), where the semantic class set is the semantic keyword list $L_s$, the color class set is color keyword list $L_c$, 
and the object class set is object keyword list $L_o$. Then, it returns the target keywords $W_s$, $W_c$, $W_o$. Last, we locate these keywords from those three dimensions and update the image recommendations accordingly. 

\textbf{(b) Text Context-Aware Exploration of Captions.} Text is also helpful in updating the caption recommendations, enabling users to explore highly related captions that resonate with the promotional goals. By using text context $T_c$, users can update the caption recommendations across various contexts, like altering the target product or changing the types of activities associated with the promotional post. To implement this feature, we leverage GPT4~\cite{openai2023gpt4} and use a prompt derived from the seed text $T_s$ and text context $T_c$ to update the caption recommendation block across three dimensions: product, activity, and advertisement. The prompt is shown below:
\lstset{
      basicstyle=\ttfamily\footnotesize,
      columns=fullflexible,
      frame=single,
      breaklines=true,
      breakindent=0pt
    }
    \begin{lstlisting}
    Prompt: Generate three promotional captions for each dimension (product, activity, advertisement) based on the given text: <T_s> and following the given prompt <T_c>. please also highlight the keywords with asterisks and keep rendering icons in each caption.
    \end{lstlisting}

\textbf{(c) Image Context-Aware Exploration of Images.} Incorporating image recommendations based on user-uploaded personal or brand logo images is a valuable feature. Users can upload images of personal interest or brand logos to enhance image recommendation results. For instance, searching for similar images that resonate with the brand's color scheme or share common objects with the brand image ensures consistency and a cohesive visual representation. To extract related keywords from the image $I_c$, we use MMDetection models leveraging with SAHI~\cite{Akyon_2022} to get the object keyword $W_o$, use ColorThief\footnote{\url{https://pypi.org/project/colorthief/}} to extract color keyword $W_c$, and use image captioning model  ofa\_image\-caption\_coco\_large\_en~\cite{wang2022ofa} and text classification model mDeBERTa-v3~\cite{laurer2022less} to get related semantic keywords $W_s$ from the semantic keyword list $L_s$. Note that to consider the exploring history, we also extract semantic keywords $W_b$ from the seed image $I_s$. The recommended images in the semantic dimension will be prioritized if they contain both keywords $W_b$ and $W_s$. Similarly, In the color and object dimension, recommendations are updated based on $W_c$ and $W_o$ keywords and we also prioritize images that include the semantic keywords $W_s$. Last, we update the image recommendation block based on the keywords we find in three dimensions.

\textbf{(d) Image Context-Aware Exploration of Captions.} Another useful contextual exploration feature is optimizing caption recommendations by incorporating personal interested images or company logo images. By uploading such images, users can re-explore caption recommendations across three contexts. This ensures that the promotional post's captions are in line with the user's preferences, the company's branding strategy, and the intended promotional message. Specifically, we use the image captioning model ofa\_image\-caption\_coco\_large\_en~\cite{wang2022ofa} to extract the information $D_i$ from the context image $I_c$. Then, we write the following prompt based on $D_i$ and the seed text $T_s$ to update the caption recommendation in three dimensions by leveraging the GPT4. The prompt is shown below:
\lstset{
      basicstyle=\ttfamily\footnotesize,
      columns=fullflexible,
      frame=single,
      breaklines=true,
      breakindent=0pt
    }
    \begin{lstlisting}
    Prompt: Generate three promotional captions for each dimension (product, activity, advertisement) based on the given text: <T_s> and following the given image prompt <D_i>. please also highlight the keywords with asterisks and keep rendering icons in each caption.
    \end{lstlisting}

\subsection{Image and Caption Fusion}

\subsubsection{\textbf{Text-based Fusion}}
In promotional post design, combining text and images, originating from text, enhances the creative process by allowing seamless integration and customization. These fusion features encompass the following key aspects, each addressing a specific aspect of text integration to optimize promotional content creation:

\textbf{(a) Text Fused with Image.} One key aspect involves fusing text prompts $T_p$ with images, where users can generate prompts to customize and regenerate images according to their promotional requirements. Similarly, we run the image captioning model ofa\_image\-caption\_coco\_large\_en ~\cite{wang2022ofa} to extract contextual information $D_i$ from the target image $I_t$, then write a prompt including $D_i$ and $T_p$ into the DALL·E~\cite{ramesh2021zeroshot} model to generate a new promotional image. The prompt we feed into the DALL·E is shown below. This fusion empowers users to tailor the visual aspects of their promotional posts, aligning with the intended message and target audience.
\lstset{
      basicstyle=\ttfamily\footnotesize,
      columns=fullflexible,
      frame=single,
      breaklines=true,
      breakindent=0pt
    }
    \begin{lstlisting}
    Prompt: <D_i> + based on the requirement: + <T_p>
    \end{lstlisting}

\textbf{(b) Text Fused with Caption.} Incorporating text seamlessly with captions is a crucial fusion feature. Users can input prompts to refine captions, making them more detailed, engaging, and aligned with the brand's messaging. Additionally, storytelling elements can be infused, augmenting the narrative of the promotional content. This fusion is implemented by concatenating the text prompt $T_p$ with the target caption $C_t$ and feeding it to the GPT-4 API~\cite{openai2023gpt4} for the generation of the new caption under the user's requirements. The caption is shown below:
\lstset{
      basicstyle=\ttfamily\footnotesize,
      columns=fullflexible,
      frame=single,
      breaklines=true,
      breakindent=0pt
    }
    \begin{lstlisting}
    Prompt: Regenerate the following promotional post caption: <C_t> based on the given text prompt: <T_p>. Please also highlight the related keywords with asterisks and keep rendering icons in the caption
    \end{lstlisting}

\subsubsection{\textbf{Image-based Fusion}}
The integration from images also forms a crucial aspect, enhancing the visual appeal and message delivery. The following elaborates various facets of image-based fusion features, focusing on customization and optimization to align with promotional objectives:

\textbf{(a) Image Fused with Image.} One fundamental aspect involves fusing a user's personal interest image $I_i$ with the target image $I_t$. This allows for customization by uploading images of personal relevance or interest, ensuring a unique and personalized touch to the promotional content. Incorporating personal images establishes a more relatable and engaging connection with the audience. Specifically, we need to extract information $D_i^i$ , $D_i^t$ from these two images using the image captioning model ofa\_image\-caption\_coco\_large\_enframework~\cite{wang2022ofa}. Then we concatenate $D_i^i$ and $D_i^t$ into a prompt, with one serving as the context and the other as the main component. This final prompt is then passed to DALL·E~\cite{ramesh2021zeroshot} for fusion. The prompt to generate a new image is shown below:
\lstset{
      basicstyle=\ttfamily\footnotesize,
      columns=fullflexible,
      frame=single,
      breaklines=true,
      breakindent=0pt
    }
    \begin{lstlisting}
    Prompt: <D_i^t> + under the context of: + <D_i^i>
    \end{lstlisting}

\textbf{(b) Image Fused with Caption.} Integrating images with captions is vital for conveying a comprehensive message. Users can upload personal or brand logo images to customize captions, making them more relevant to the company's image or aligning with specific areas of interest. This fusion provides an opportunity to infuse brand identity and captivate the audience with intriguing visual-textual combinations. Similarly, we use the image captioning model ofa\_image\-caption\_coco\_large\_en ~\cite{wang2022ofa} to extract the information $D_i$ from the context image $I_i$. Then, we write the following prompt based on $D_i$ and the target caption $C_t$ to generate a new caption under the image context. The prompt to refine the caption is shown below:
\lstset{
      basicstyle=\ttfamily\footnotesize,
      columns=fullflexible,
      frame=single,
      breaklines=true,
      breakindent=0pt
    }
    \begin{lstlisting}
    Prompt: Regenerate the following promotional caption <C_t> based on the given image context: <D_i>. Please also highlight the related keywords with asterisks and keep rendering icons in the caption.
    \end{lstlisting}

%% file: 5_UserEvaluation.tex
\section{User Evaluation}

\begin{figure}[!tb]
  \centering
  \includegraphics[width=\linewidth]{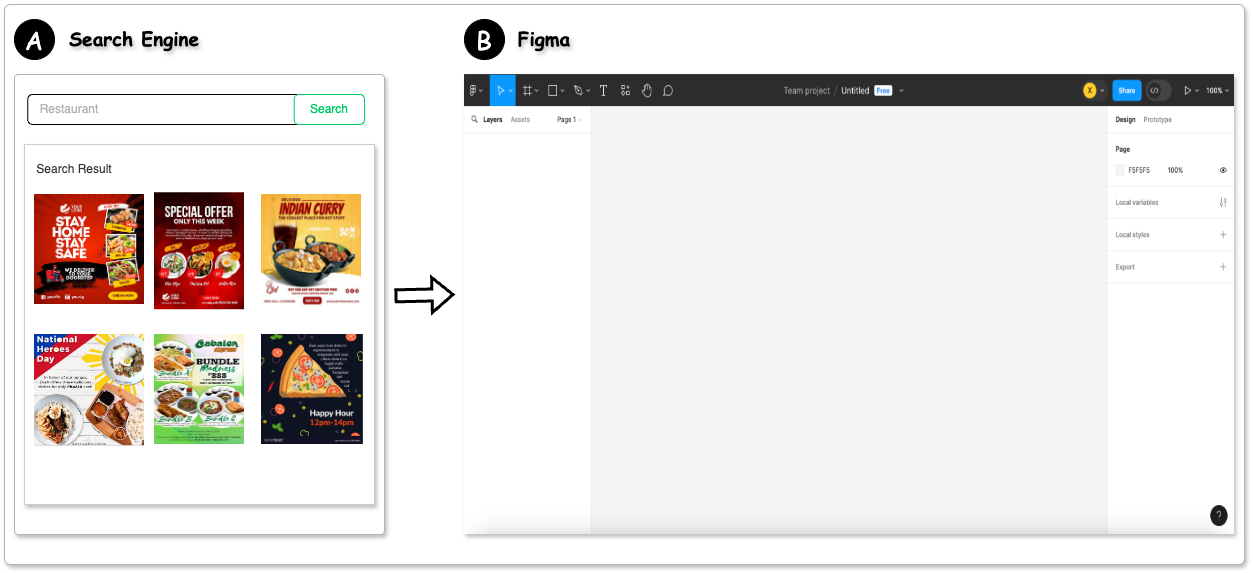} 
  \vspace{-8mm}
  \caption{The Baseline system includes two parts: (A) search engine and (B) Figma, which resembles the current workflow with industry standard tools.} 
\label{fig:baseline}
\Description{} 
\end{figure}

We conducted a within-subjects controlled experiment comparing \sysname{} with a Baseline including Google Search for ideation and  Figma for post design (\autoref{fig:baseline}). Since promotional post design is a complex process, we structured a within-subjects study to mitigate the potential impact of individual differences in this creative process. According to insights from our formative study, Google Search stands as the most commonly utilized platform for seeking inspiration, and Figma is also commonly used by users, making it a suitable baseline system for promotional post design. 
This baseline featured a Google Search-like interface with essential functions such as image searching (utilizing the same database) and image saving capabilities. Users can then use Figma to customize the image and write the caption to create a promotional post. 

\subsection{Participants}

We recruited 12 participants (7 women and 5 men, age: \mean{27.3}, \sd{3.6}) via social media and mailing lists. The majority of them (10 out of 12) need to engage with graphic design in their daily tasks, although they are not proficient designers. With a pre-study questionnaire, they were considered novice designers based on their self-reported design experiences on a 5-point Likert Scale, including expertise in promotional post design (\md{1}, \sd{0.42}) as well as familiarity
with color theories (\md{1.5}, \sd{0.78}), and Figma (\md{1}, \sd{0.56}), where 1 indicates less expertise/ familiarity. 
Thus, they are representatives of the typical target users for \sysname{} (e.g., marketers, product managers, UI designers, and small business owners) who want to easily and quickly produce a few promotional post designs for broadcasting events or products.

\subsection{Tasks and Design}

We employed a within-subjects design for the study. 
We designed three tasks to compare \sysname{} with the Baseline on multiple aspects of creativity and assistance in promotional post design tasks. 
All tasks require users to design a promotional post from the provided topic which is the most common case for novices to not start from scratch. 
We set a time limit (60 minutes) for each task to simulate a timing design scenario. 
This also allowed the study length to be reasonable in a within-subject design. 

\textbf{Task 1} focuses on a product promotion scenario. Given a product topic ``Fruit Juice", users need to design relevant images to comply with the topic. They also need to generate the related captions to be compatible with the images and make the final post design harmonic overall.

\textbf{Task 2} focuses on the scenario of activity promotion. Provided with an activity theme: ``Basketball Game", participants need to design the promotional image to achieve a design matching the theme. Participants should also generate reasonable captions to express a particular design need and suitably describe the image. There should be at least two versions of the final product with different colors to accommodate different sentiments or occasions. 

\textbf{Task 3} aims to test the brand-based promotion design scenario. Given a brand image and topic: ``Ramen", participants need to design an image compatible with this brand image on the canvas, which includes the color of the image, and the object of the image. In addition, the final design should be aesthetically pleasing.

\subsection{Procedure}

During the study, participants complete the above three tasks using both of the study systems (i.e., \sysname{} and Baseline), one after another. The order and combination of study data and tools were counterbalanced. For each condition, participants were first introduced to the study system (i.e., Baseline or \sysname{}). 
During the \sysname{} condition, a short tutorial video demonstrating its basic functionality was shown to users. Once users were comfortable using the tool, tasks were presented sequentially with the final deliverables for each task. Post task completion, each participant completed a questionnaire related to their experience, comparing the Baseline and the \sysname{}. The questionnaire included the Creative Support Index (CSI) assessment~\cite{cherry2014quantifying} regarding users’ experience in exploration, expressiveness, enjoyment, workload, etc. Besides, we conducted a short semi-structured interview with each participant to gain their qualitative feedback. The entire study lasted approximately 60 minutes for each participant, and they were remunerated with \$20.

\section{Results}\label{user evaluation}

\begin{figure*}[tb]
  \centering
  \includegraphics[width=\linewidth]{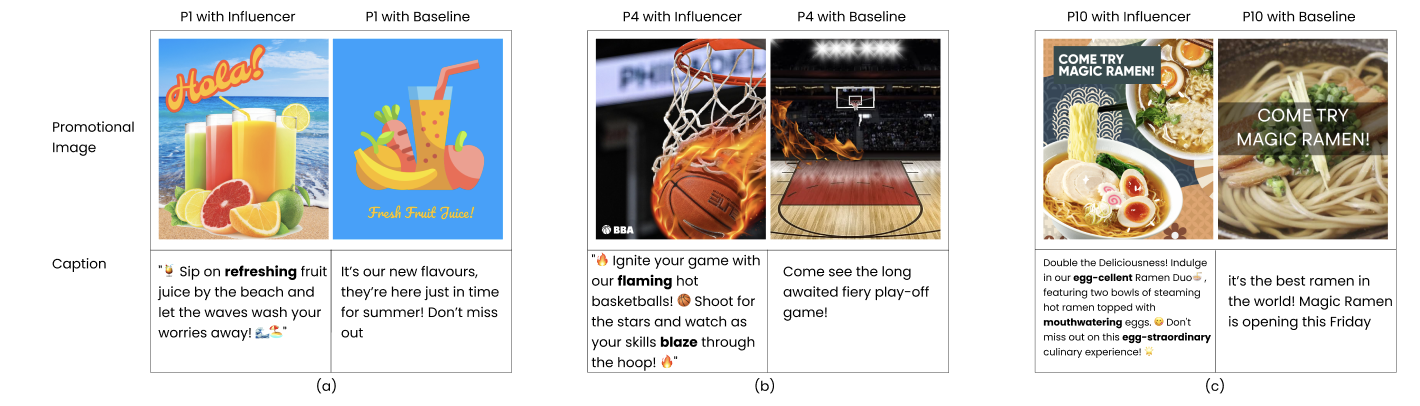} 
  \vspace{-9mm}
  \caption{Examples of six drafted posts generated by three participants (P1, P4, P10) with \sysname{} and Baseline system in three tasks: (a) a fruit juice promotional post for a small grocery store, (b) a post for promoting basketball game activity, and (c) a ramen advertisement for a Japanese restaurant. }
\label{fig:example}
\Description{}
\end{figure*}

\subsection{Quantitative Results} 
In the following, we report our results on task completion rate and participants' ratings for \sysname{} and Baseline.

\textbf{Completion rate.} With \sysname{}, all participants claimed they had completed all the required promotional posts within the given time for the three tasks. With the Baseline, 91.67\% (11/12) of participants completed the post design in Task 1, but only 75.00\% (9/12) of the post design in Task 2 and 83.3\% (10/12) in Task 3 were fully completed. This indicates that \sysname{} has the potential to accelerate design processes when designers need to edit images or quickly generate alternative designs. Here, we present six representative drafts among 72 drafts generated with \sysname{} and the baseline system in \autoref{fig:example}.

\begin{figure}[tb!]
  \centering
  \includegraphics[width=\linewidth]{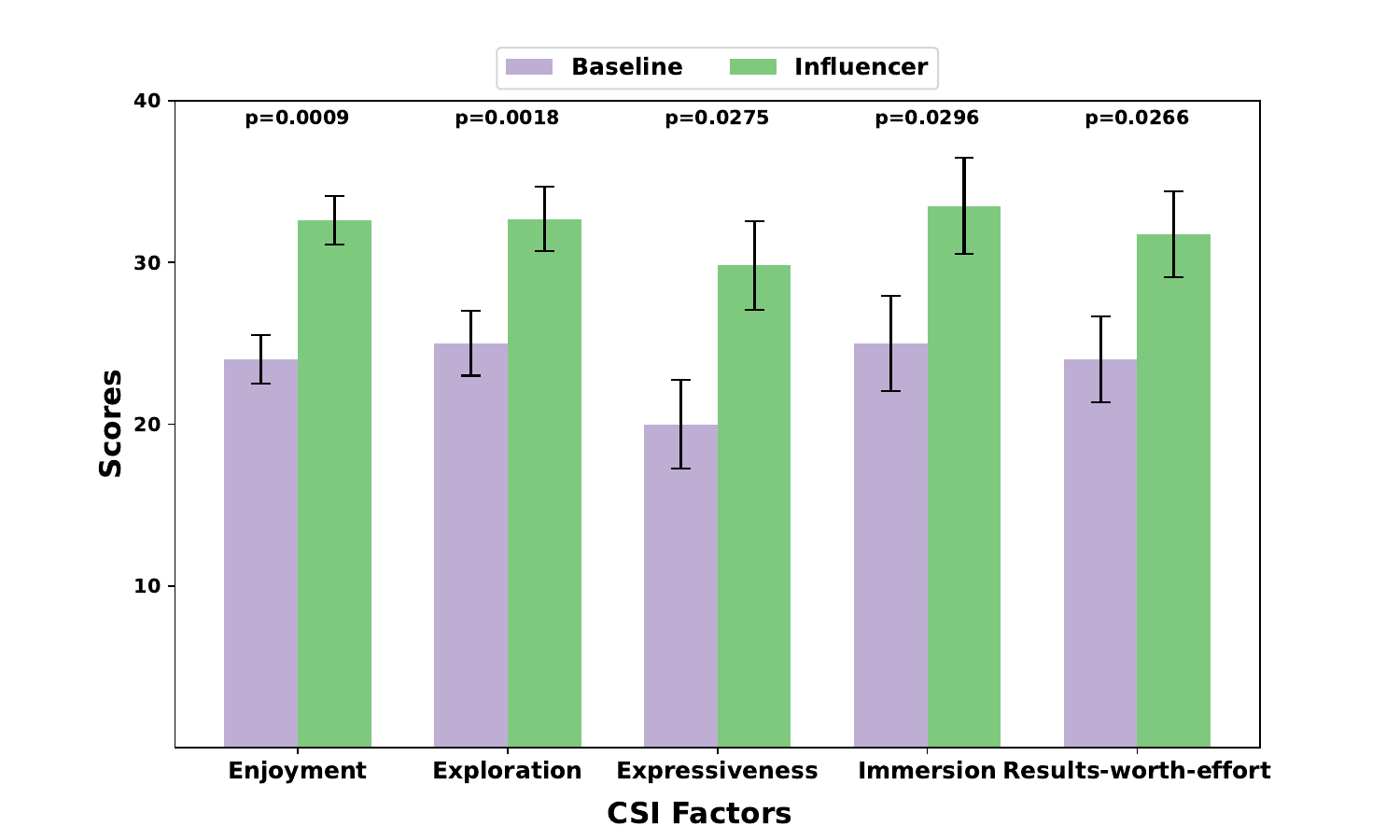} 
  \vspace{-8mm}
  \caption{Results of Creative Support Index (CSI) for \sysname{} and Baseline (the higher the better) on the factors of enjoyment, exploration, expressiveness, immersion, and results worth effort. } \label{fig:csi}
  \Description{}
\end{figure}

\textbf{Design Experience.} We utilized the Creativity Support Index (CSI)~\cite{cherry2014quantifying} to measure the degree of creativity support for \sysname{} and the Baseline in the study. Participants rated five creativity support factors with scores on a Likert scale from 0 (strongly disagree) to 10 (strongly agree). \autoref{fig:csi} shows the individual CSI score for each factor, i.e., enjoyment, exploration, expressiveness, immersion, and results-worth-effort. Overall, \sysname{} achieved a CSI score of 76.86 (\sd{13.52}), which is higher than the Baseline with a score of 61.5 (\sd{16.64}). 
The paired t-tests show that the overall CSI score of \sysname{} is significantly higher than that of Baseline ($t = 4.136, p = .0007$). It was found that \sysname{} generated statistically significant improvements in enjoyment ($t = 3.362, p = .0009$), exploration ($t = 3.445, p = .0018$), expressiveness ($t = 2.762, p = .0275$), immersion ($t = 2.614, p = .0296$), and results worth effort ($t = 3.184, p = .0266$). The results indicate that participants enjoyed their overall experience with \sysname{}. \sysname{} supported graphic design explorations effectively, enhanced users' expressiveness during the creative process and increased their satisfaction with their design outcomes.

 \textbf{Usefulness of Functions}. We also evaluated different functionalities within \sysname{} to see which ones are comparatively more useful with a 5-point Likert scale: mind-map based Visual Representation (\mean{4.1}, \sd{.71}), Materials Fusion (\mean{3.64}, \sd{1.24}), Post Generation (\mean{3.9}, \sd{.84}), Caption Recommendation (\mean{3.42}, \sd{.65}), Semantic Recommendation for Image (\mean{3.88}, \sd{1.24}), Color Recommendation for Image (\mean{2.43}, \sd{1.57}), Object Recommendation for Image (\mean{3.18}, \sd{1.16}), Regenerate Image (\mean{2.36}, \sd{.89}), Mask Edit Image (\mean{2.85}, \sd{1.26}). 
 As we can see, mind-map based Visual Representation, Post generation, and Dragging Interaction were recognized as the most useful functions in \sysname{}. However, we also find that some functions have deviations, including materials fusion, semantics/color/object exploration, and mask edit images. According to our observations, this kind of disagreement in usefulness evaluation reflects different user preferences with different post design habits and the limitation of image generation models (see \autoref{sec:discussion} for more discussion). 

 \subsection{Qualitative Results}

 In general, participants appreciated various novel functions of \sysname{}. We summarize their feedback from the semi-structured interviews based on the following themes. 

\textbf{General preference.} Overall, 10 out of 12 participants stated that they would like to use \sysname{} as their promotional post design tool over the baseline system. Two other participants preferred to combine the two systems to do image searches and get image and caption recommendations but did image customization on another interface. 

\textbf{Image/ caption recommendation}  Participants thought the overall quality of the image-searching results met their expectations. Eight noted that the images retrieved were notably creative and useful compared to their prior experiences using a search engine (P1-2, P4, P10-12). They also mentioned the usability of three-dimensional recommendation for image and caption and stated \qt{I would appreciate using this function when lacking initial ideas on post design. The recommendations fit my design goal well. The three-dimensional recommendation helps me find more inspiration with just a few seed images.} (P6). In addition, participants mentioned that even though there are some recommended images are not suitable for the topic, it is still useful since \sysname{} already saves them more time in designing a final version of the post design (P8). Further, \qt{Based on those recommendations, the post design process becomes easier and reduces my mental effort required for brainstorming.} (P7). Meanwhile, three participants (P4, P10, P11) recognized the three recommendation directions on caption based on different contexts: \qt{I can quickly change my post captions to fit different scenes, making it much easier than starting from scratch.} (P9). The feedback indicates that the multi-dimensional image and caption recommendation is useful to facilitate the promotional post design with desired content.

\textbf{Context-Aware Exploration quality}  
All participants found that the context-aware recommendation is helpful. The utility of context-aware exploration was highlighted when exploring the images and captions based on the given materials: \textit{e.g.,} \qt{it swiftly guides users to find post images and captions that align with their given brand logo or product message in terms of color, semantics, and objects"} (P1-3), \qt{context awareness on caption exploration is nice since I can get a bunch of captions at a time matching my requirement of linguistic preferences or product relevance"} (P5). Moreover, the logical interaction of context-aware exploration was also recognized by 3 participants(P6, P7, P10): \qt{It is very user-friendly to allow me to update the recommendation stuff by dragging the uploaded images or created text prompt to the target recommendation block.} Another participant also reported: \qt{It tells me why this recommendation block is updated, so I can track and know what to focus on.}

\textbf{Flexible fusion of various images and captions.} This is a unique experience reported by eight participants. They found it to be very useful when they wish to directly customize an image or caption by writing a prompt or uploading the image they possess (P1-2, P8, P10-12). By flexibly fusing the image or caption, they could \qt{freely customize the material based on my preference and iteratively brainstorm}(P12). P1 also mentioned \qt{It is very helpful when I want to rewrite a caption incorporating the image I possess since it directly helped me integrate the information from the images into the previous captions smoothly.} P4 echoed, \qt{I feel it convenient to design the captions or images based on my personal preference by using different prompts or images. It helps me get multiple alternatives to compare later.} 

\textbf{Mind-map layout to track thinking path and record design alternatives.}  Four participants found it very useful to trace back their thinking path (P6-8, P10). By enabling tracking design history, they can \qt{check their design process, freely fuse different materials, and review multiple design alternatives.} P7 also added that \qt{It records my thoughts and helps me track my thinking process. Moreover, I indeed feel secure with this feature, and enables me to explore freely.} Another participant also mentioned different colors for dragging features helped them clearly differentiate the exploration process and customize the process (P11). He recognized the \sysname{} \qt{showing how the initial thoughts have been refined and making it easier for users or team members to follow the thought process and rationale behind decisions} P12 added, \qt{I can personally upload an image or write a prompt to customize the image or caption exploration path, which is very flexible and helps me find the related image efficiently.}

\subsection{Expert Assessment}

\begin{figure}[tb!]
  \centering
  \includegraphics[width=\linewidth]{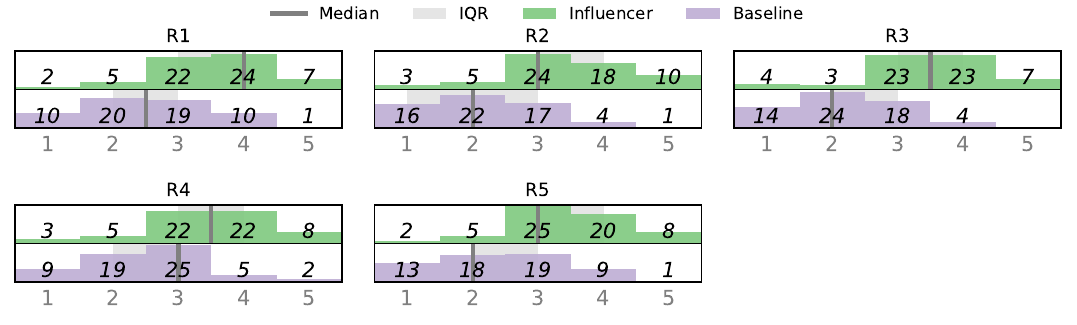} 
  \vspace{-8mm}
  \caption{Experts' ratings on the quality of participant-generated promotional posts on a 5-point Likert scale (the higher the better) for five aspects: overall satisfaction (R1), clarity of content (R2), audience engagement (R3), content consistency (R4), and aesthetics (R5).  } \label{fig:expert}
  \Description{}
\end{figure}

We recruited three expert designers to help us rate and analyze participant promotional post designs. We also invited them to share their views regarding \sysname{} afterward. The experts' backgrounds are as follows: 1) an assistant professor, a man, specializing in graphic design with a PhD degree, with 15 years of design; 2) a visual designer in the industry, a woman, focusing on advertising design, holding a master's degree in design theory, with eight years of design experience; 3) an art teacher, a woman, teaching art design in high school and usually create design contents for the school's social media, with ten years of design. We denote them as EA1-3, respectively.

\textbf{Post Quality Assessment}. While the CSI assessment in our questionnaire indicates participants' self-experience of the created promotional post, these do not reflect how the slides are received by the audience. To assess the quality of the promotional posts created with \sysname{}, we invited the three experts to rate the post drafts from 12 participants. For a fair comparison, we removed the trials from the two conditions if the participants did not complete that task with either of the tools. Each expert rated the promotional posts by participants on a 5-point Likert Item and in random order. The results of their ratings on five different aspects are shown in \autoref{fig:expert}. 
The Wilcoxon signed-rank tests found that the outcomes of \sysname{} had significantly better ratings than those of the Baseline for all aspects: overall satisfaction ($t = 3.245, p = .0060$), clarity of content ($t = 2.864, p = .0405$), audience engagement ($t = 2.693, p = .0398$), content consistency ($t = 3.843, p = .0270$), and aesthetics ($t = 3.325, p = .0318$). The results indicate that \sysname{} could effectively facilitate users to generate better quality promotional posts, more consistent with the given topic, and more engaging, aesthetic, and clear for audiences. 

\textbf{Design of \sysname{}}. The three-dimensional recommendation framework for both images and captions was recognized as reasonable by all our experts. EA1 raised that \qt{This brainstorming framework is helpful for novices to quickly learn basic design elements, save time on searching, and generate high-quality drafts, especially on post captions.} EA2 reported that even this tool is \qt{intriguing to interact with and useful especially when we have hands-on materials to design,} and he would like to give it a try. Moreover, EA3 gave high praise for the flexible fusion of various materials and also for tracking their thinking path with the mind-map on canvas. She suggested that all materials need to be allowed to be fused or linked to freely generate new examples: \qt{using mind-map on canvas is an easy way to achieve this action.}

%% file: 6_Discussion.tex
\section{Discussion} \label{sec:discussion}

This paper focuses on promotional post design, enabling everyday users to express their creativity such as in micro-entrepreneurship. 
However, the design of promotional posts requires high requirements on design skills as it involves blending engaging visuals with effective messages to produce appealing content, particularly challenging for design novices. To address these challenges, we have designed and evaluated the \sysname{}, an AI-infused tool designed to make the process of designing promotional posts simpler for those without professional design skills. 
\sysname{} integrates four design modules: ideation, context-aware exploration, flexible fusion, and a mind-map layout for tracking the thinking path. 
As indicated by the results, using \sysname{} could streamline the promotional post design process, enabling users to craft engaging and effective promotional content with relative ease. 
These experiences have been further supported and concertized by the participants' detailed descriptions in the qualitative findings. Beyond presenting a novel design case and contextually confirming the benefits of our tool, 
we discuss the extracted underlying patterns and implications for future research and design below.

\subsection{Examples and Context as Key Components in Design}

Studies have been conducted to explore the idea of a recommendation framework on how users could get inspiration by single visual elements~\cite{kim2017thoughts, peterson2019aspects, jahanian2017colors}. Kang \etal also uses a multi-dimensional recommendation method to enrich richer image search~\cite{kang2021metamap}. However, little has been done to understand and leverage the context-aware exploration when specific design requirements are present. Our work has surfaced that context-aware exploration can refine the exploration process beyond simply a linear recommendation, especially when users face complex design scenarios that require tailored solutions. 
Our findings from Section \ref{user evaluation} reveal distinct views of three dimensions in the context-aware exploration: 
\qt{Semantic dimension is easy to use and understand.} (P2 and P8). \qt{Color is useful in most of cases especially when I upload a brand image and want to find related images in a similar color schema.} (P10), \qt{Object dimension can help me find images having similar objects which are helpful when I need to design a product-type promotional post.} (P11). But P6 and P7 expressed concern about color and objects in one of the tasks they did. They mention the image recommendation on color and object dimensions sometimes is slightly inaccurate. The reason may be the inaccurate color extraction and misclassification of objects caused by the limitation of the object detection model, leading to colors and objects that do not match the users intend to use in their uploaded brand image. 
To better support this approach in designing, one design expert (E2) suggested that 
adding a color picker to let users choose preferred colors, and then recommend related color palettes may be helpful to improve image recommendation accuracy. For object dimension, enhancing the model and training it with more object classes on a large-scale dataset would be an addition to the context-aware exploration.



\subsection{Flexible Fusion and Generation with Personalization} 
As uncovered by our empirical data, another promising opportunity of \sysname{} is that it could provide an interactive way to arrange and fuse elements and ease the design process by the drag-and-drop feature on canvas. 
These new experiences in promotional post design have not adequately exhibited in existing creative tools. They usually offer template-based designs, where users select from pre-defined layouts and styles, limiting the scope for personalized interaction and creativity~\cite{10.1145/2502081.2502116, qiang2017learning, Tyagi_2022}. The future design could build upon this experiential interaction and make the design process more immersive. For example, in 3D design, when a user drags one 3D model close to another, the system could automatically align and merge these geometric shapes, creating a unified and complex entity. Further, as lighting elements are integrated into the scene, their effects could intelligently merge, adapting to create a cohesive and harmonized lighting atmosphere that enhances the overall visual impact of the 3D space.
Besides, based on our user studies, we learned that users may have different preferences on the design styles of post images and captions. Therefore, we may leverage data analytics and machine learning to analyze user behavior, preferences, and past design choices, offering tailored design recommendations on images and captions. This is also an interesting avenue to consider in the future development of more personalized recommendation models in design. 

\subsection{Trade-off between Automation and Autonomy} 
To facilitate the creation of promotional posts by design novices, \sysname{} employs a nuanced recommendation system that carefully balances guidance with creative freedom.
The variety of examples collected in the findings illustrates that a multi-dimensional recommendation framework on images and captions enriches the design process by offering a diverse array of visual and textual elements, allowing users to explore design elements using different dimensions or concepts. Further, users were able to tailor the design process to the user's specific brand narrative and aesthetic preferences with context-aware exploration. 
In our study, participants could quickly customize the images and captions in the post design with little manual effort. The results indicate that \sysname{} outperforms the baseline on various design tasks. However, this customization is limited by the performance of the Generative Image AI models. 
For example, 
\qt{The image-based fusion feature sometimes doesn't work as I expected. Newly generated images didn't change the style of my dragged image.} (P12). 
Fine-tuning the Generative Image AI model in a post image dataset could improve the visual quality of generated post images. 
Allowing users to edit and polish the images later would also be a helpful addition to the system, such as adding text to the image, adjusting image saturation, and other basic features like those in Photoshop. 
In summary, this suggests that an ideal design system should balance automation and autonomy to provide a wealth of inspiration with recommendation, generation, and customization abilities.

\subsection{Limitations and Future Work} 
This work has several limitations which we plan to address in the future. First, even though our system can auto-layout each block, the layout is still stiff when users create a large amount of alternatives. Users can only use the mouse to manually drag each overlapped block to satisfy their layout requirement. To enable a more flexible mind-map layout, we could adopt a self-defined mind-map editor for the personalized organization. In this way, users can save the useful block based on their needs. 

Second, \sysname{} recommends images based on three dimensions: semantic, object, and color. However, the current visual algorithm has limitations in accurately identifying highly related images within complex images. In addition, when users upload their own images and attempt to update the image recommendation block in our system, existing vision models are unable to correctly extract the semantic meaning of the uploaded image. To improve the accuracy and integrity of recommendation algorithms, we would like to expand our dataset and experiment with a more state-of-art vision model in the future. 

Third, since there are many kinds of information including texts, photos, typography design, and layout that can inspire users, our three-dimensional exploration would benefit by some extension. To adapt our system to more creative scenarios (e.g., banner design, photography, etc), we can include more images in the dataset or recommend images in more dimensions (e.g., typography, mood, style). Also to help professional designers, we can incorporate \sysname{} with Photoshop to perform more complex tasks or leverage advanced image editing features from Photoshop to fulfill more advanced user needs.

Fourth, while users can modify various design elements, such as images, captions, and color schemes, the flexibility of fully customizing layouts and structures remains somewhat constrained. As part of our future endeavors, we envision enhancing the system's capabilities so users can design completely customized layouts incorporating diverse design elements precisely based on their preferences. Users can integrate features that facilitate seamless collaboration among multiple users participating in promotional post creation. Future work may try to streamline and enrich the design process, promoting more dynamic and efficient teamwork in creating compelling promotional content by enabling real-time collaboration and feedback.

%% file: 8_Conclusion.tex
\section{Conclusion}

In this paper, we have introduced \sysname{}, a tool designed to support novice designers such as freelance creators, marketers, and product managers with promotional post design. By seamlessly integrating captivating images and well-crafted captions, \sysname{} addresses the challenges faced by design novices in generating attention-grabbing content. Our system offers a mindmap-like layout for ideation and incorporates multidimensional AI-powered exploration and customization of related images and captions to revolutionize the design process. Additionally, \sysname{} allows flexible fusion of various design elements. Through a comprehensive evaluation, including controlled experiments and expert assessment, we have demonstrated that \sysname{} significantly enhances the ideation and design process. It enhances the effectiveness of promotional post design in different task scenarios by fostering engaging interactions, diverse explorations, and trackable thought processes.